\documentclass[journal,twocolumn,10pt]{IEEEtran}

\usepackage{amsmath,amsfonts}
\usepackage[ruled,linesnumbered]{algorithm2e}
\usepackage[caption=false,font=footnotesize,labelfont=rm,textfont=rm]{subfig}
\usepackage{makecell}
\usepackage{stfloats}
\usepackage{ragged2e}
\usepackage{url}
\usepackage{verbatim}
\usepackage{graphicx}
\usepackage{balance}
\usepackage{cite}
\usepackage{multirow}
\usepackage{enumitem}
\usepackage{tikz}
\usetikzlibrary{shapes, arrows.meta, positioning}
\usepackage{booktabs}
\usepackage{lipsum}
\usepackage{amssymb}
\usepackage{etoolbox}
\usepackage[colorlinks,linkcolor=blue,anchorcolor=blue,citecolor=blue]{hyperref}

\makeatletter
\patchcmd{\@makecaption}
{\\}
{.\ }
{}
{}

\newcommand{\myrefeq}[1]{(\ref{#1})}
\newcommand{\myreffig}[1]{Fig. \ref{#1}}
\newcommand{\myreftable}[1]{Table \ref{#1}}
\newcommand{\myrefalgo}[1]{Algorithm \ref{#1}}

\title{Towards Intelligent Low-Altitude Wireless Network Deployment: Differentiable Channel Knowledge Map Construction and Trajectory Design}
\author{
	Le Zhao, Zesong Fei, \textit{Senior Member, IEEE}, Wenge Shi, Xinyi Wang, \textit{Member, IEEE}, Jingxuan Huang,\\ Jihao Luo, and Yong Zeng, \textit{Fellow, IEEE}
	\thanks{
		Le Zhao, Zesong Fei, Wenge Shi, Xinyi Wang, Jingxuan Huang, and Jihao Luo are with the School of Information and Electronics, Beijing Institute of Technology, Beijing 100086, China.
		Wenge Shi is with the ZhongGuanCun (ZGC) Institute of Ubiquitous-X Innovation and Applications, Beijing 100088, China.
		Yong Zeng is with the National Mobile Communication Research Laboratory, School of Information Science and Engineering, Southeast University, Nanjing 211189, China. Yong Zeng is also with Purple Mountain Laboratories, Nanjing 211111, China.
		(e-mail: tobin$\_$bit@icloud.com, feizesong@bit.edu.cn, bit$\_$wangxy@163.com, shiwenge@zgc-xnet.com, jxhbit@gmail.com, jihaoluo$\_$bit@icloud.com, yong$\_$zeng@seu.edu.cn)

		Part of this paper \cite{zhao2025diffkan} was presented at the 2025 International Conference on Wireless Communications and Signal Processing (WCSP 2025), Chongqing, China.
	}
	\vspace{-0.1in}
}

\begin{document}
	\maketitle
	\begin{abstract}
		Channel knowledge map (CKM) has emerged as a promising technique to leverage prior propagation knowledge in low-altitude wireless networks (LAWNs), yet state-of-the-art grid-based CKM construction methods  struggle to support efficient LAWN deployment due to their lack of differentiability with respect to continuous locations of unmanned aerial vehicles (UAVs). To overcome this limitation, we propose a differentiable CKM-triggered trajectory optimization framework for LAWNs.
		Firstly, we propose a location-oriented CKM construction method that directly maps continuous spatial coordinates to channel gain. In particular, a shared convolutional neural network (CNN) is employed to encode high-level environmental features from conditional inputs. These features are then sampled based on location information to form a fused regressor—conditional multilayer perceptron (c-MLP) or conditional Kolmogorov-Arnold network (cKAN)—for channel gain prediction.
		We further propose a joint power, bandwidth, and trajectory optimization (JPBTO) method for multi-UAV systems, with the constructed differentiable CKM employed to evaluate the communication performance. The formulated non-convex problem is solved via alternating optimization and successive convex approximation. Numerical results show that the proposed framework enables location-aware differentiability of the CKM, while achieving higher accuracy than the methods without environmental features. Furthermore, the proposed CKM-JPBTO achieves a significantly higher minimum throughput than conventional statistical channel model-based JPBTO.
	\end{abstract}


	\vspace*{-0.1in}
	\section{Introduction}
	\IEEEPARstart{L}{ow}-altitude wireless networks (LAWNs), primarily enabled by cellular-connected unmanned aerial vehicles (UAVs), have emerged as a key scenario for the sixth generation (6G) wireless networks due to their high mobility and low deployment cost \cite{WC_Zeng_cellularUAV_2019, commag_fei}. Through seamlessly interacting with existing cellular infrastructures, LAWNs are expected to support low-altitude economy applications such as aerial filming \cite{flyvideo}, target tracking \cite{targettrack}, and aerial transportation \cite{Telikani_UAV_transportation_ST_2025}. These missions require stable communication links with terrestrial base stations (BSs) to ensure reliable remote control and real-time data streaming \cite{UAVadZ}.
	
	In practice, the air-to-ground (A2G) communication links is significantly influenced by signal attenuation due to distance and blockages caused by obstacles \cite{A2Gchannel_TCOM_JiangHao_2020}. Consequently, optimizing the trajectory of UAV nodes has drawn significant research attention to enhance overall network performance. Existing studies often assume perfect channel state information or known channel models, and employ differentiable functions to map UAV positions to link quality, thereby facilitating gradient-based trajectory optimization. These approaches predominantly rely on location-aware differentiable models, such as distance-dependent path loss with probabilistic line-of-sight (LoS) conditions. As a pioneering contribution, \cite{ZengYong_TWC_UAV_trajectory_2017} investigated energy-efficient trajectory design for communication-aware UAV operation. This work was later extended to multi-UAV systems in \cite{ZengYong_WQQ_TWC_multi_UAV_traj_2018}, which addressed the problem of maximizing the minimum user rate and inspired numerous subsequent studies on UAV trajectory design. As a step further, \cite{ZhangRui_TWC_2020_pLoS} studied trajectory design under a probabilistic LoS channel model in UAV-assisted sensor data collection networks, while \cite{Rician} derived the optimal hovering altitude for UAVs under Rician fading to minimize outage probability. In \cite{GWO_UAV_delpoy2024}, the authors considered the coverage-energy trade-off in multi-UAV systems, a grey wolf optimizer (GWO) \cite{MIRJALILI201446}-based multi-objective framwork was proposed to jointly optimize the number, positions, and velocities of UAVs for coverage and energy efficiency. {While these pioneering models provide mathematically tractable expressions for efficient optimization, they primarily capture general trends. Overlooking fine-grained building blockages, they often struggle to adapt to complex, site-specific propagation environments.}
	
	The concept of channel knowledge map (CKM) has recently been introduced to address these challenges \cite{ZengYong_2021_IEEEWC, ZengYong_2024_IEEE_ST}. By constructing a channel map that reflects the signal propagation mechanisms under specific environmental topologies, CKMs provide location-aware prior knowledge, significantly enhancing wireless performance in LAWNs and vehicle-to-everything communications. Traditional CKM construction methods—including K-nearest neighbors (KNN) \cite{KNN_2009_TIP}, Kriging \cite{Krige_2017_TVT}, and spatial correlation-based model fitting \cite{Data4CKMConstruc_2024_TWC}—rely heavily on channel gain measurements sampling. These approaches, therefore, suffer from high computational complexity in large-scale environments and limited accuracy due to their dependence on statistical assumptions.
	Subsequent efforts have turned to convolutional neural networks (CNNs), such as RadioUNet \cite{Levie_2021_TWC}, which leverage spatial correlations to improve reconstruction accuracy. Graph neural networks (GNNs) \cite{li2023graph} further extended this by modeling environmental relationships via graph structures, enhancing generalization across diverse settings. More recently, generative models—including generative adversarial networks \cite{ChenJunting_2024_ICC} and diffusion-based methods \cite{zeng2024generative, wang2024radiodiff, Le3DRadiodiff2025wcl, Trans_3DRadioDiff_2025}—have been employed to reconstruct high-resolution CKMs from sparse measurements, demonstrating strong robustness to environmental variations. Despite these advances in accuracy, {these grid-based CKMs introduce a paradigm mismatch for UAV trajectory planning. Since UAV flight is a continuous physical process, discrete CKMs are fundamentally non-differentiable with respect to locations, preventing seamless integration with rigorous gradient-based trajectory optimization.}
	
	In response to the non-differentiability of existing CKMs, several studies have adopted methods that do not require position differentiation—such as discrete position optimization to optimize trajectories of cellular-connected UAVs. For example, \cite{ABS_deployment_tcom_2024} discretized UAV deployment and selected optimal relay positions using propagation-based CKMs, which avoids unknown derivatives but cannot guarantee global optimality over continuous space. Following a similar discretization philosophy, \cite{SINRCKM} constructed SINR-based CKMs and formulated trajectory optimization as a shortest-path searching problem in graph theory. In parallel, \cite{DQL} and \cite{DQL18} applied deep reinforcement learning (DRL) to optimize UAV trajectories for sum rate improvement by learning a direction-value function. Furthermore, \cite{CKM_DQL} further used in-flight channel measurements to jointly train the CKM and trajectory optimizer, and evaluated performance via predicted CKM to refine the DRL model. These methods, however, {involve inherent trade-offs. Discretizing the continuous flight space \cite{ABS_deployment_tcom_2024, SINRCKM, DQL, DQL18} scales poorly due to the curse of dimensionality. Meanwhile, although \cite{CKM_DQL} jointly updates the models, DRL typically struggles to strictly guarantee physical kinematic constraints. Conversely, gradient-based solvers remain highly desirable, offering deterministic convergence and robust constraint handling. Thus, bridging discrete CKMs with continuous gradient solvers is critical.}
	
	To address the need for location-aware differentiable CKMs, several coordinate-input methods have been proposed to learn mappings from continuous locations to channel information using prior models such as free-space path loss and shadowing residuals. Techniques include expectation maximization \cite{CKMEM}, maximum likelihood estimation \cite{MLE}, deep Gaussian processes \cite{DGP}, and multilayer perceptrons (MLPs). Leveraging inherent differentiability, MLP-based CKMs have been incorporated with convex optimization based trajectory design in non-terrestrial networks \cite{lyMLPCKM2025tvt}. Yet, {mapping pure coordinates to complex channels lacking the spatial inductive biases of CNNs requires massive datasets to capture 3D building topologies. Therefore, exploring a hybrid paradigm that merges the rich spatial feature extraction of grid-based CKMs with the analytical differentiability of coordinate-based models remains a vital frontier.}

	This paper addresses these limitations by proposing a novel multi-UAV communication optimization framework built upon a differentiable CKM constructed via Kolmogorov–Arnold networks (KANs) \cite{kan_paper2024}. In particular, KANs utilize B-spline basis functions to learn complex functional mappings with high interpretability and accuracy. Building upon the KAN-based CKM introduced in our prior work \cite{zhao2025diffkan}, which provides continuous and differentiable channel gain predictions with respect to BS position, we further integrate environmental conditions via a convolutional neural network (CNN) encoder into the regressor for channel gain, proposing an model-data dual-accelerated CKM construction mechanism termed conditional KAN (cKAN). Based on the cKAN-based CKM, we then investigate CKM-based joint optimization of power, bandwidth, and trajectory (CKM-JPBTO) to maximize the minimum communication rate among UAVs.
	The main contributions of this paper are summarized as follows:
	\begin{itemize}
		\item We propose an encoder-regressor architecture to construct location-aware and differentiable CKMs, where a CNN encoder extracts environmental features to condition a coordinate-based regressor. This framework supports both conditional MLP (cMLP) and conditional KAN (cKAN) implementations. Notably, by exploiting the efficient B-spline activation of KANs, the developed cKAN achieves superior reconstruction fidelity with significantly higher parameter efficiency compared to cMLP, offering a lightweight solution for aerial platforms.
		\item We establish a differentiable optimization bridge that integrates the constructed CKM into a multi-UAV joint power, bandwidth, and trajectory optimization (JPBTO) framework. We leverage the chain-rule differentiability of the proposed cKAN/cMLP to compute explicit gradients of channel gain with respect to continuous UAV locations. This mechanism enables the optimizer to theoretically exploit specific environmental reflection paths via gradient ascent, ensuring efficient solutions via alternating optimization (AO) and successive convex approximation (SCA).
		\item Through extensive simulations, we validate the high accuracy of the cKAN-based CKM construction and the effectiveness of the proposed CKM-JPBTO algorithm. By comparing the performance with that of joint optimization based on statistical channel models (SC-JPBTO), we demonstrate the necessity of CKM for environment-aware trajectory and transmission design.
	\end{itemize}
%

	The remainder of this paper is organized as follows. Section II presents the system model and problem formulation for the multi-UAV communication system. Section III elaborates on the construction of the differentiable CKM using the proposed cKAN. Section IV details the proposed CKM-JPBTO. Section V provides numerical results to validate the accuracy of the cKAN-based CKM and the effectiveness of the proposed optimization framework, and insights into CKM accuracy impacts. Section VI concludes the paper.
	
	\begin{figure}[!t]
		\centering
		\includegraphics[width=0.85\columnwidth]{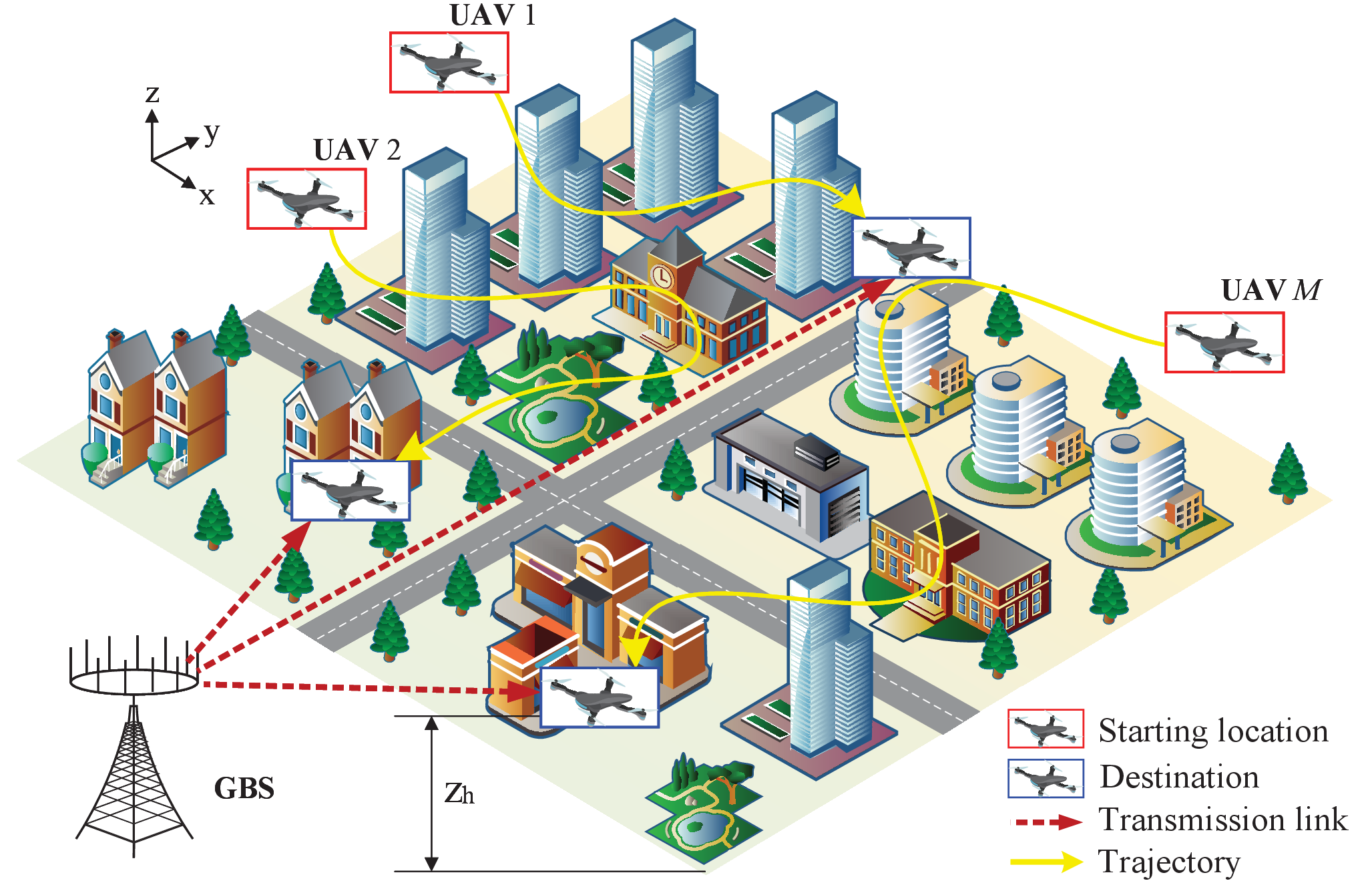} 
		\caption{Illustration of the multi-UAV communication system in an urban scenario.}
		\label{fig:min_rate_vs_M}
		\vspace{-0.1in}
	\end{figure}
	
	\vspace{-0.1in}
	\section{System Model and Problem Formulation}

	\subsection{System Model}
	We consider a downlink communication system depicted in \myreffig{fig:min_rate_vs_M}, where a terrestrial BS simultaneously transmits data to multiple cellular-connected UAVs under frequency division multiple access.
	The mission area is modeled as a 3D cubic space with dimensions $\{W_1, W_2, H\} \in \mathbb{N}^+$ (corresponding to length, width, and height of the mission area, respectively), containing distributed buildings blocking the line-of-sight (LoS) link.
	Considering that the UAVs fly at a fixed altitude $z_h$, we use $\mathbf{q}_m[n] = [x_m[n], y_m[n]]^T \in \mathcal{D}$ to denote the location of UAV $m \in \mathcal{M}$, where $\mathcal{M} \triangleq \{1,\dots,M\}$. Given a horizontal spatial resolution $\Delta \kappa$, we define $N_{W_1}=W_1/\Delta \kappa, N_{W_2}=W_2/\Delta \kappa$; hence, $\mathcal{D}$ is discretized with $\tilde{q} = [\tilde{x} \Delta \kappa, \tilde{y} \Delta \kappa]^T \in \tilde{\mathcal{D}}$, where $\tilde{x} \in \{1, \dots, N_{W_1}\}$, $\tilde{y} \in \{1, \dots, N_{W_2}\}$, and $\tilde{\mathcal{D}} \in \mathbb{R}^{N_{W_1} \times N_{W_2}}$. We also define the environment topology as $\mathbf{E} \in \mathbb{R}^{N_{W_1} \times N_{W_2}}$, with ${B}^{\rm env}_{(\tilde{x},\tilde{y})} \in \mathbf{E}$ denoting the building height at grid location $(\tilde{x},\tilde{y})$, and the buildings coverage range in $\mathcal{D}$ is defined as $\mathbf{q}_{\text{obs}, k} \in \mathcal{O}, k \in \{1,\dots,K\}$, where $K$ is the number of buildings. UAVs need to keep a certain distance, $D_{\rm \min}$, from the buildings to ensure flight safety.

	In each time slot, the downlink system is served using the total bandwidth $B_{\max}$ and the maximum transmit power $P_{\max}$. All UAVs are required to maneuver from some given starting locations to various destinations within the given flight period $T$, with the starting location and destination of UAV $m$ denoted as $\mathbf{q}_{m}^{\rm start}$ and $\mathbf{q}_{m}^{\rm end}$, respectively. The total flight period $T$ is discretized into $N$ equal time slots, with the duration of each being $\tau =T/N$, and the set of time slots is defined as $\mathcal{N} \triangleq \{1,\dots,N\}$.

	Denote the location of BS as $\mathbf{q}_{\rm BS} = [x_{\rm BS},y_{\rm BS}]$, where $\mathbf{q}_{\rm BS} \in \mathcal{D}$, with its height being $z_{\rm BS}$. The overlapped received signal reflected or diffracted by buildings at the $m$-th UAV location $\mathbf{q}_{m}$ can be expressed as a superposition of $L$ multipath components, expressed as
	\begin{align}\label{eq:channel_definition}
		Y(\mathbf{q}_m, \mathbf{q}_{\rm BS}) = \sqrt{p_m} \sum_{l=1}^{L} \bar{h}_l(\mathbf{q}_m) \tilde{h}_l(\mathbf{q}_m) X + Z,
	\end{align}
	where $X \sim \mathcal{CN}(0,1)$ is the transmitted signal, $p_m$ is transmit power, $Z \sim \mathcal{CN}(0, \sigma^2)$ is Gaussian noise, where $\sigma^2$ is the noise power.

	Ray-tracing techniques \cite{imai2017survey} can be employed to model the channel from $\mathbf{q}_{\rm BS}$ to $\mathbf{q}_m$ as expressed in \myrefeq{eq:channel_definition}.
	Specifically, the large-scale fading $\bar{h}_l(\mathbf{q}) = \sqrt{\frac{G_{t,l} G_{r,l}}{{\rm PL}_l(\mathbf{q}_m)}}$ captures deterministic attenuation, with path loss expressed as
	\begin{align}
		{\rm PL}_{l}(\mathbf{q}_m) = \prod_{i=1}^{I_R} |\Gamma_i|^{-2} \prod_{i=1}^{I_D} |\Lambda_i|^{-2} \left( \frac{4\pi d_l(\mathbf{q}_m) f}{c} \right)^2,
	\end{align}
	where $d_l(\mathbf{q}_m)$ is the propagation distance, $f$ is carrier frequency, $c$ is light speed, $I_R$ is the number of reflections (each with material- and angle-dependent coefficient $\Gamma_i$, $0 < |\Gamma_i| < 1$), and $I_D$ is the number of diffractions (each with coefficient $\Lambda_i$, $0 < |\Lambda_i| < 1$). This formulation links ${\rm PL}_l$ to interaction counts and types, escalating with more reflections or diffractions.

	For small-scale fading, since ray-tracing determines all paths geometrically, $\tilde{h}_l(\mathbf{q}_m)$ can be modeled deterministically for each distinct path as $\tilde{h}_l^{\rm d}(\mathbf{q}_m) = e^{j \phi_l(\mathbf{q}_m)}$, where $\phi_l(\mathbf{q}_m) = \frac{2\pi d_l(\mathbf{q}_m)}{\lambda} + \sum \phi_{i}^{(\cdot)}$ incorporates phase shifts from distance and interactions, $(\cdot)$ can be $R$ or $D$, denotes the phase shifts from reflection and diffraction. The overall fading emerges from coherent summation across paths. To account for unmodeled uncertainties like minor scatterers or positioning errors, an optional random perturbation can be added as multiplicative Gaussian noise $\tilde{h}_l^{\rm r}(\mathbf{q}_m) \sim \mathcal{CN}(e^{j \phi_l(\mathbf{q}_m)}, \sigma_h^2)$, and the aggregate small-scale fading of $l$-th path can be expressed as $\tilde{h}_l(\mathbf{q}_m) = \tilde{h}_l^{\rm d}(\mathbf{q}_m)\tilde{h}_l^{\rm r}(\mathbf{q}_m)$.

	The CKM with respect to a BS maps physical locations $\mathbf{q} \in \tilde{\mathcal{D}}$ to the expected aggregate channel power gain.
	Denote $\mathcal{H}(\mathbf{q}_m)$ as the ground-truth mapping function of channel gain  towards the BS $\mathbf{q}_{\rm BS}$ and $m$-th UAV at location $\mathbf{q}_m$, it can be expressed as
	\begin{align}
		\mathcal{H}(\mathbf{q}_m) = \mathbb{E} \left[ \left| \sum_{l=1}^{L} \bar{h}_l(\mathbf{q}_m) \tilde{h}_l(\mathbf{q}_m) \right|^2 \right].
	\end{align}

	Under this setup, the downlink transmission rate for the $m$-th UAV at the $n$-th time slot is given as
	\begin{align}\label{eq:rate_expression}
		R_m[n] = \alpha_m[n] B_{\max} \log_2\left(1 + \frac{p_m[n] \mathcal{H}(\mathbf{q}_m[n])}{N_0 \alpha_m[n] B_{\max}}\right),
	\end{align}
	where $\alpha_m[n] \in [0, 1]$ is the normalized bandwidth allocation coefficient,  $N_0$ represents the noise power spectral density.
	Over $N$ time slots, the $m$-th UAV's average achievable rate ${R}^a_{m}$ can be calculated by
	\begin{align}
		\label{eq:avergae_rate}
		{R}^a_{m}=\frac{1}{N}\sum_{n=1}^{N}R_{m}[n].
	\end{align}

	Since obtaining the ground-truth CKM is generally unavailable, especially in the continuously dense set $\mathcal{D}$. To reduce the measurement costs, mentioned works \cite{KNN_2009_TIP, Krige_2017_TVT, Data4CKMConstruc_2024_TWC, Levie_2021_TWC, li2023graph, ChenJunting_2024_ICC, zeng2024generative, wang2024radiodiff, Le3DRadiodiff2025wcl, Trans_3DRadioDiff_2025, CKMEM, MLE, DGP, lyMLPCKM2025tvt} firstly discretize the location set $\mathcal{D}$ to obtain a discreted locations $\mathbf{q} \in \tilde{\mathcal{D}}$ and a grid-based ground-truth CKM $\mathbf{\Psi} \in \mathbb{R}^{N_{W_1} \times N_{W_2}}$. Thereafter, the predicted CKM $\hat{\mathbf{\Psi}}$ by image processing netwoks, based on the sparse channel power gains sampled from $\tilde{\mathcal{D}}$ denoted as location $\mathbf{q} \in \mathcal{D}^{\rm s}$, which also is a subset of $\mathcal{D}$. As mentioned before, these methods lack location-aware differentiable and cannot be extended to dense or continuous location sets.

	In this paper, we aim to learn by neural network (NN) to construct a more accurate mapping function, with differentiablity to location input. As for the achievable received rate in \myrefeq{eq:rate_expression} and \myrefeq{eq:avergae_rate}, we replace the $\mathcal{H}(\mathbf{q})$, as well as $R_m$, and $R_m^a$, with the predicted mapping function $\hat{\mathcal{H}}(\mathbf{q}) = \mathcal{F}(\mathbf{q})$, as well as $\hat{R}_m$ and $\hat{R}_m^a$.
	The achieved rate can be expressed as
	\begin{align}\label{eq:approx_rate}
		\hat{R}_m[n] = \alpha_m[n] B_{\max} \log_2\left(1 + \frac{p_m[n] \mathcal{F}(\mathbf{q}_m[n])}{N_0 \alpha_m[n] B_{\max}}\right),
	\end{align}
	and the approximated average achievable rate is $\hat{R}^a_{m}=\frac{1}{N}\sum_{n=1}^{N}\hat{R}_{m}[n]$.
	In this system, the allocated bandwidth for the $m$-th UAV at time slot $n$ is $b_m[n] = \alpha_m[n] B_{\max}$, and allocated transmit power of $m$-th UAV is $p_m[n]$.
	The key notations used in this paper are summarized in Table \ref{tab:notation}.

	\begin{table}[!t]
		\caption{System notations}
		\label{tab:notation}
		\centering
		\renewcommand{\arraystretch}{1.05}
		\setlength{\tabcolsep}{4mm}
		\begin{tabular}{l|l}
			\hline
			\textbf{Symbol} & \textbf{Description} \\
			\hline
			$M$ & Number of UAVs \\
			$N$ & Number of time slots \\
			$B_{\max}$ & Total bandwidth \\
			$P_{\max}$ & Maximum transmit power of terrestrial BS\\
			$L$ & Number of multipath components \\
			$\mathbf{q}_m[n]$ & Position of $m$-th UAV at slot $n$ \\
			$\bar{h}_l(\cdot)$ & Large-scale channel fading of $l$-th path \\
			$\tilde{h}_l(\cdot)$ & Small-scale channel fading of $l$-th path  \\
			$\alpha_m[n]$ & Bandwidth coefficient for $m$-th UAV at slot $n$ \\
			$b_m[n]$ & Allocated bandwidth for $m$-th UAV at slot $n$\\
			$p_m[n]$ & Transmit power of $m$-th UAV at slot $n$ \\
			$N_0$ & Noise power spectral density \\
			$\Gamma_i$ & Fading coefficient of reflections \\
			$\Lambda_i$ & Fading coefficient of diffractions \\
			$\mathcal{H}(\cdot)$ & Ground-truth channel gain mapping function\\
			$\hat{\mathcal{H}}(\cdot), \mathcal{F}(\cdot)$ & Approximated channel gain mapping function\\
			$\mathbf{\Psi}$ & Ground-truth CKM \\
			$\hat{\mathbf{\Psi}}$ & Predicted CKM \\
			$\mathcal{D}$ & Set of locations of mission area \\
			$\Delta \kappa$ & horizontal spatial resolution \\
			$\mathcal{D}^{\rm s}$ & Sampling sub-set of mission area \\
			$\tilde{\mathcal{D}}$ & Discrete set of $\mathcal{D}$ with resolution $\Delta \kappa$ \\
			$V_{\text{max}}$ & Maximum UAV flight speed \\
			$\tau$ & Time slot duration \\
			$\mathbf{q}_{\text{start}}, \mathbf{q}_{\text{end}}$ & UAV start and end positions \\
			$\mathbf{q}_{{\rm obs},k}, D_{\min}$ & Position and safety distance of $k$-th obstacle \\
			$\mathbf{E}$ & environment topology \\
			$R_{\min}$ & Minimum rate threshold of UAVs \\
			\hline
		\end{tabular}
		\vspace*{-0.2cm}
	\end{table}

	\subsection{Problem Formulation}

	In this work, we aim to maximize the system's minimum throughput over the entire period among all UAVs by jointly optimizing the trajectories $\mathcal{Q} = \{\mathbf{q}_m[n], m \in \mathcal{M}, n\in \mathcal{N}\}$, bandwidth allocation coefficients $\mathcal{A} = \{\alpha_m[n], m \in \mathcal{M}, n\in \mathcal{N}\}$, and transmit power $\mathcal{P} = p_m[n], m \in \mathcal{M}, n\in \mathcal{N}$, while satisfying the constraints of power budget, UAV mobility, and the flying safety. The optimization problem can be formulated as follows:
	\begin{align} \label{prob0}
		{\rm (P0)}:	\quad \max_{\mathcal{Q,A,P}}\quad  &  \min_m \left( \frac{1}{N} \sum_{n=1}^N R_m[n] \right) \\
		{\rm s.t.} \quad & \epsilon \leq \alpha_m[n] \leq 1 \,\,\, \forall m, n \tag{\ref{prob0}a} \\
		& \sum_{m=1}^M\alpha_m[n] = 1,\,  \sum_{m=1}^M p_m[n] = P_{\text{max}}, \,\, \forall n \tag{\ref{prob0}b} \\
		& \Vert\mathbf{q}_m[n+1] - \mathbf{q}_m[n]\Vert_2 \leq V_{\text{max}} \cdot \tau, \,\, \forall m, n \tag{\ref{prob0}c} \\
		& \mathbf{q}_m[1] = \mathbf{q}_m^{\text{start}}, \,\, \mathbf{q}_m[N] = \mathbf{q}_m^{\text{end}}, \,\, \forall m \tag{\ref{prob0}d} \\
		& \mathbf{q}_m[n] \in \mathcal{D}, \,\, \forall m, n \tag{\ref{prob0}e} \\
		& \Vert\mathbf{q}_m[n] - \mathbf{q}_{{\rm obs},k}\Vert_2 \geq D_{\min}, \,\, \forall m, n, k \tag{\ref{prob0}f} \\
		& R_m[n] \geq R_{\text{min}}, \,\, \forall m, n \tag{\ref{prob0}g},
	\end{align}
	where $\epsilon$ is an arbitrarily small positive constant to prevent $\alpha_m[n]$ from causing numerical instability.
	Here, (\ref{prob0}a)-(\ref{prob0}c) are the bandwidth, power and speed constraints of UAVs, where $V_{\max}$ m/s in (\ref{prob0}c) is the maximum flying speed of UAVs. (\ref{prob0}d)-(\ref{prob0}e) are the UAV maneuvering constraints of UAVs. (\ref{prob0}f) is the obstacle avoidance constraint. (\ref{prob0}g) is the minimum rate constraint.
	The problem \myrefeq{prob0} is a non-convex, multi-variable coupled optimization problem due to the non-convexity of the rate expression with respect to trajectory and bandwidth, and the non-convexity of the obstacle avoidance constraint. Existing image-feature-based CKMs, proposed in the \cite{Levie_2021_TWC, li2023graph, ChenJunting_2024_ICC, zeng2024generative, wang2024radiodiff, Le3DRadiodiff2025wcl, Trans_3DRadioDiff_2025}, are non-differentiable with respect to the UAV locations $\{\mathbf{q}_m\}_{1\leq m\leq M}$, and thus, cannot be applied in the UAV trajectory design via typical gradient-based optimization algorithms. Therefore, a location-aware differentiable CKM construction mechanism is proposed, and the alternating optimization algorithm is introduced to address the coupled variables.

	\section{Location-aware differentiable CKM construction}
	In this section, we aim to construct a differentiable CKM with respect to BS.
	Specifically, we first propose an environment-feature encoder to obtain the environment feature maps. Subsequently, the fused features, combined with location coordinates and environment features, are input to the KAN-based regressor to enable a rapid and cost-efficient way  fpr location-aware differentiable CKM construction.

	\begin{figure*}[!t]
		\centering
		\includegraphics[width=1.85\columnwidth]{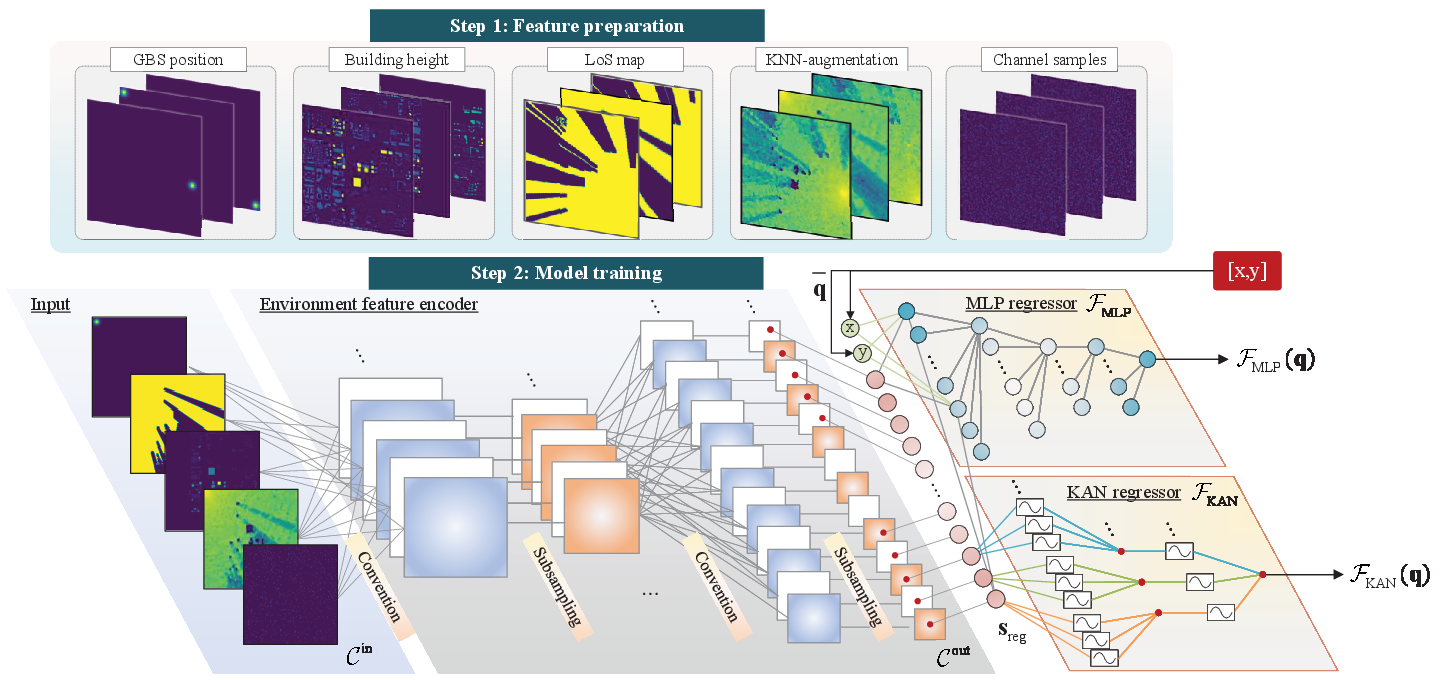}
		\caption{The illustration of the proposed model-data dual-accelerated CKM construction architecture. The CNN is employed to encode the multi-channel input features into high-dimensional feature maps. Normalized location coordinate and condition input vector gridded from the encoded feature maps are combined and input into the MLP and KAN joint regressor to predict the channel gain.}
		\label{fig:structure}
		\vspace{-0.1in}
	\end{figure*}

	\vspace{-0.1in}
	\subsection{Feature Preparation}

	To enhance the accuracy of the constructed CKM, we employ a model-data dual-accelerated mechanism. Specifically, to enhance the model's ability to capture wireless propagation dynamics, a set of multi-channel environmental feature maps, derived from ray-tracing simulations or real-world measurements, is employed as input, providing rich prior information about the environment. The employed input features include BS position, building height map, and sampled channel gain map studied in the previous works. Furthermore, we also incorporate the LoS map and KNN augmentation into the multi-dimensional input features.

	Details on the LoS map and KNN augmentation based input features are introduced as follows:
	\begin{enumerate}
		\item We define the LoS map as $\mathbf{L} \in \{0,1\}^{N_{W_1} \times N_{W_2}}$, with the element $L_{(\tilde{x},\tilde{y})}$ indicating whether the LoS path between the BS and the UAV at location $\tilde{\mathbf{q}}=[\tilde{x}, \tilde{y}]^T \in \tilde{\mathcal{D}}$ exists. To compute this, we evaluate the path from $\mathbf{q}_{\rm BS}$ to $\tilde{\mathbf{q}}$ at height $z_{\rm h}$: if any building height ${B}^{\rm env}_{(x,y)} \in \mathbf{E}$ along the path exceeds the path height (via linear interpolation), ${L}_{(\tilde{x},\tilde{y})} = 0$; otherwise, ${L}_{(\tilde{x},\tilde{y})} = 1$.
		\item As mentioned in our conference paper \cite{zhao2025diffkan}, we use KNN-interpolation by the sampled channel gain measurements $\mathbf{\Psi}^{\rm s} \in \mathbb{R}^{N_{W_1} \times N_{W_2}}$ to enhance the training dataset. Note that the augmented dataset provides denser samples, and improves the performance, the training cost is also increased. In this paper, the KNN-interpolated map $\hat{\mathbf{\Psi}}^{\rm KNN} \in \mathbb{R}^{N_{W_1} \times N_{W_2}}$ is incorporated into the input features, and the grid-based CNN feature processing directly merges the KNN-augmentation into the high-dimensional feature map.
	\end{enumerate}

	During the model training, the input features are all normalized and stacked along the channel dimension. Consequently, the input tensor to the subsequent environmental feature encoder is denoted as $\mathcal{C}^{\rm in} \in \mathbb{R}^{d^{\rm in} \times H^{\rm in} \times W^{\rm in}}$, where $d^{\rm in} = 5$, $H^{\rm in} = N_{W_1}$, and $W^{\rm in} = N_{W_2}$.

	\vspace{-0.1in}
	\subsection{Model Architecture}

	The model architecture fuses the prepared conditions via a CNN backbone, followed by location-aware feature gridding and regression to predict channel gains, constructing the CKM as $\mathcal{H}(\mathbf{q}) = \mathcal{F}(\mathbf{q};\mathcal{C})$, where $\mathcal{F}(\cdot)$ is the learned mapping.

	\subsubsection{Environment Feature Encoder}

	The CNN backbone processes the multi-channel input condition tensor $\mathcal{C}^{\rm in}$ to extract a high-dimensional feature map. The encoder captures multi-scale spatial dependencies inherent in wireless propagation, such as obstacle-induced shadowing, LoS conditions, and interpolation priors from the KNN-interpolation map, by learning hierarchical representations from the diverse input priors.
	Mathematically, $\mathcal{F}_{\rm CNN}(\cdot)$ is a sequence of convolutional operations, non-linear activations, residual blocks, and pooling layers designed to abstract environmental features while preserving spatial locality, which is denoted as $\mathcal{C}^{\rm out} \in \mathbb{R}^{d^{\rm out} \times H^{\rm out} \times W^{\rm out}}$, where $d^{\rm out}$, $H^{\rm out}$, and $W^{\rm out}$ are set during the model designing.


	\subsubsection{Location-aware Regressor}

	To enable continuous prediction of the channel gain with respect to location $\mathbf{q}$, we employ a differentiable location-aware mechanism that samples features from the CNN output $\mathcal{C}^{\rm in}$ and jointly regresses the channel gain with MLP or KAN. This hybrid design ensures that channel gain $\mathcal{H}(\mathbf{q}) = \mathcal{F}(\mathbf{q}; \mathcal{C})$ is both accurate and differentiable, overcoming the discrete, non-differentiable nature of CNN-based discerted CKM construction. We define that $\bar{\mathbf{q}} = \mathcal{N}(\mathbf{q}) = [\bar{x}, \bar{y}]^T$ is the normalized location coordinate, which can be calculated as
		\begin{align}\label{eq:normalize_q}
			\bar{x} = \frac{x-X^{\min}}{X^{\max}-X^{\min}}, \quad \bar{y} = \frac{y-Y^{\min}}{Y^{\max}-Y^{\min}},
		\end{align}
		where $X_{\min}, Y_{\min}$ are the minimun $x, y$ coordinate of location set $\mathcal{D}$, $X_{\max}, Y_{\max}$ are the maximum $x, y$ coordinate of $\mathcal{D}$.

	Location-specific features are sampled via bilinear interpolation, expressed as
	\begin{align}
		\label{eq:loc_spec_feature_sample}
		\mathbf{s} = \mathcal{I}(\mathcal{C}^{\rm out}, \bar{\mathbf{q}}') \in \mathbb{R}^{d^{\rm out} \times 1}, \quad \bar{\mathbf{q}}' = 2\bar{\mathbf{q}} - 1
	\end{align}
	where $\mathcal{I}$ is a grid sampling function that computes a smooth weighted average from the four nearest grid points in each channel of CNN output feature maps $\mathcal{C}^{\rm out}$. For a extend normalized coordinate $\bar{\mathbf{q}}' = 2\bar{\mathbf{q}} - 1$, let $(i, j)$ be the integer grid cell and $(\mu, \nu)$ the fractional offsets (where $0 \leq \mu, \nu < 1$). The sampled value for channel $d$ is
	\begin{align}
		s_{(d)} = & (1-\mu)(1-\nu) \mathcal{C}^{\rm out}_{(d,i,j)} + \mu(1-\nu) \mathcal{C}^{\rm out}_{(d,i+1,j)} \\ & \qquad \qquad + (1-\mu)\nu \mathcal{C}^{\rm out}_{(d,i,j+1)} + \mu\nu \mathcal{C}^{\rm out}_{(d,i+1,j+1)}, \notag
	\end{align}
	where $s_{(d)}$ is the $d$-th element of $\mathbf{s}$.
	The sampled features $\mathbf{s}$ are then concatenated with the explicit location coordinates as
	\begin{align}\label{eq:fusion_vector_input}
		\mathbf{s}_{\rm reg} = \bar{\mathbf{q}} \oplus \mathbf{s} \in \mathbb{R}^{1 \times (2 + d^{\rm out})},
	\end{align}
	where $\oplus$ forming a joint input that conditions the regression on both location information and fused environmental features. This concatenation is crucial for capturing location-dependent variations from the input condition tensor $\mathcal{C}^{\rm out}$, enhancing precision over location-only inputs.

	For the baseline MLP regressor, we employ a feedforward network with fixed activations to predict the channel gain.
	With the combined location-specific features, the regressor maps $\mathbf{s}_{\rm reg}$ to the predicted channel gain can be expressed as
	\begin{align}\label{eq:MLP_regression}
		\hat{\mathcal{H}}(\bar{\mathbf{q}}) & = \mathcal{F}_{\rm MLP}(\bar{\mathbf{q}}; \mathcal{C}^{\rm in}) \\
		& = \mathcal{R}_{\rm MLP}(\mathbf{s}_{\rm reg}) = \sigma (\mathbf{W}_{L} \cdots \sigma (\mathbf{W}_{1} \mathbf{s}_{\rm reg} + \mathbf{b}_1) \cdots + \mathbf{b}_L), \notag
	\end{align}
	where $\mathcal{R}$ represnts the regression network, $l \in  \{1, \dots, L\}$, $\mathbf{W}_l \in \mathbb{R}^{d_{l-1} \times d_l}$ and $\mathbf{b}_l \in \mathbb{R}^ {d_l \times 1}$ are learnable weights and biases, $\sigma$ is ReLU activity function, $d_l$ is the layer dimension. This structure approximates the channel gain via linear transformations and piecewise non-linearities, but its fixed activations limit expressivity in complex propagation scenarios.

	As a step further, we propose a KAN regressor to enhance channel gain prediction, building on our prior conference work \cite{zhao2025diffkan}. Unlike MLP, which relies on fixed ReLU activations, KAN employs adaptive univariate transformations, making it particularly suited to capture intricate wireless channel variations driven by location and environmental features.
	The KAN regressor maps the joint input $\mathbf{s}_{\rm reg} = \bar{\mathbf{q}} \oplus \mathbf{s} \in \mathbb{R}^{1 \times (2 + d^{\rm out})}$ to the predicted channel gain. Inspired by the Kolmogorov-Arnold theorem \cite{kan_paper2024}, KAN decomposes multivariate functions into compositions of univariate mappings. For a KAN with $L_{\max}$ layers, the prediction is expressed as:
	\begin{align}
		\hat{\mathcal{H}}(\bar{\mathbf{q}}) & = \mathcal{F}_{\rm KAN}(\bar{\mathbf{q}}; \mathcal{C}^{\rm in}) \\
		& = \mathcal{R}_{\rm KAN}(\mathbf{s}_{\rm reg}) = \phi_L \circ \phi_{L-1} \circ \cdots \circ \phi_1(\mathbf{s}_{\rm reg}), \notag
	\end{align}
	where $\circ$ denotes function composition, and each layer $\phi_l$ transforms an input vector $\mathbf{v}^{(l-1)} \in \mathbb{R}^{n_l}$ (with $\mathbf{v}^{(0)} = \mathbf{s}_{\rm reg}$) into an output vector $\mathbf{v}^{(l)} \in \mathbb{R}^{n_{l+1}}$. The $j$-th output component of layer $l$ is computed as:
	\begin{align}
		v_j^{(l)} = \sum_{i=1}^{n_l} \psi_{l,i,j}(v_i^{(l-1)}), \quad \text{for } j = 1, \dots, n_{l+1},
	\end{align}
	where $\psi_{l, i, j}$ is a learnable univariate function parameterized as a B-spline of order $k=4$ over $g=10$ grid intervals, expressed as
	\begin{align}
		\psi_{l,i,j}(v) = \sum_{m=0}^{g+k-2} c_{l,i,j,m} B_{m,k}\left( \frac{v - t_0}{t_g - t_0} \cdot g \right),
	\end{align}
	with $c_{l,i,j,m}$ as trainable coefficients and $B_{m,k}$ as the B-spline basis functions. These are defined recursively via the Cox-de Boor formula:
	\begin{align}
		B_{m,1}(v) = \begin{cases}
			1 & \text{if } t_m \leq v < t_{m+1}, \\
			0 & \text{otherwise},
		\end{cases}
	\end{align}
	and for $k>1$, we have
	\begin{align}
		& B_{m,k}(v) = \\
		& \quad \frac{v - t_m}{t_{m+k-1} - t_m} B_{m,k-1}(v) + \frac{t_{m+k} - v}{t_{m+k} - t_{m+1}} B_{m+1,k-1}(v), \notag
	\end{align}
	where $\mathbf{t} = [t_0, \dots, t_{g+2k-2}]$ is a non-decreasing knot sequence, clamped at boundaries to ensure endpoint interpolation. This spline-based design enables the KAN to capture non-linear dependencies.
	The overall cKAN-based CKM construction is shown in \myrefalgo{algo:train}, and the training loss funcion is designed as
	\begin{align}
		\label{loss_function}
		\mathcal{L} = \mathbb{E}\left\{ \left[ \hat{\mathcal{H}}(\bar{\mathbf{q}}) - \mathcal{H}(\mathbf{q}) \right]^2 \right\},
	\end{align}
	where $\bar{\mathbf{q}} = \mathcal{N}(\mathbf{q})$ is the nomalized coordinates via \myrefeq{eq:normalize_q}.

	\begin{algorithm}[!t]
		\caption{cKAN-based CKM Construction}
		\label{algo:train}
		\SetAlgoLined
		\KwIn{Multi-channel input features $\mathcal{C}^{\rm in} \in \mathbb{R}^{d^{\rm in} \times H^{\rm in} \times W^{\rm in}}$, normalized location coordinates $\bar{\mathbf{q}} \in [0,1]^2$}
		\KwOut{Predicted channel gain $\hat{\mathcal{H}}(\bar{\mathbf{q}})$}
		Extract high-dimensional feature map: $\mathcal{C}^{\rm out} = \mathcal{F}_{\rm CNN}(\mathcal{C}^{\rm in})$ \;
		Sample location-specific features via bilinear interpolation: $\mathbf{s} = \mathcal{I}(\mathcal{C}^{\rm out}, 2\bar{\mathbf{q}} - 1)$\;
		Form joint regression input: $\mathbf{s}{\rm reg} = \bar{\mathbf{q}} \oplus \mathbf{s}$\;
		Regress channel gain using KAN: $\hat{\mathcal{H}}(\bar{\mathbf{q}} = \mathcal{R}_{\rm KAN}(\mathbf{s}_{\rm reg}) = \mathcal{F}_{\rm KAN}(\bar{\mathbf{q}}; \mathcal{C}^{\rm in})$;
	\end{algorithm}
	
	\subsection{End-to-end Gradient Propagation}
	
	The core innovation of our CKM is its differentiability with respect to the location $\mathbf{q}=[x, y]^T$, enabling gradient-based optimization for tasks like UAV trajectory design.
	To make full use of the activate function, the location coordinate input is normalized to the values within $[0,1]$ as $\bar{\mathbf{q}} = [\bar{x}, \bar{y}]^T$\footnote{Please note that this normalization is a linear transformation, which has no effect on differentiability. Therefore, for the sake of simplification, the normalized $\bar{\mathbf{q}} = [\bar{x}, \bar{y}]^T$ in Section IV is denoted by $\mathbf{q} = [x,y]^T$.}.
	Unlike prior image-based methods, which produce discrete, non-differentiable maps, or coordinate-only regressions which lack environmental context, our hybrid model integrates CNN-extracted features with explicit location inputs, ensuring both accuracy and {piecewise-smooth} gradients.
	The channel gain $\hat{\mathcal{H}}(\bar{\mathbf{q}}) = \mathcal{F}(\bar{\mathbf{q}}; \mathcal{C}^{\rm in})$ is calculated as
	\begin{align}
		\hat{\mathcal{H}}(\bar{\mathbf{q}}) & = \mathcal{F}_{\rm KAN}(\bar{\mathbf{q}}; \mathcal{C}^{\rm in}) = \mathcal{R}_{\rm KAN}\left(\mathbf{s}_{\rm reg}\right) \\ & = \mathcal{R}_{\rm KAN}\left(\bar{\mathbf{q}} \oplus \mathcal{I}\left(\mathcal{F}_{\rm CNN}\left(\mathcal{C}^{\rm in}\right), 2\bar{\mathbf{q}} - 1\right)\right), \notag
	\end{align}
	where $\mathcal{F}_{\rm CNN}$ is the CNN feature extractor, $\mathcal{I}(\cdot)$ is bilinear interpolation, and $\mathcal{R}_{\rm KAN}$ is the KAN regressor. The gradient of the CKM with respect to ${\bar{\mathbf{q}}}$ is
	\begin{align}
		\nabla_{\bar{\mathbf{q}}} \hat{\mathcal{H}}(\bar{\mathbf{q}}) = \left[ \frac{\partial \mathcal{F}_{\rm KAN}(\bar{\mathbf{q}}, \mathcal{C}^{\rm in})}{\partial \bar{x}}, \frac{\partial \mathcal{F}_{\rm KAN}(\bar{\mathbf{q}}, \mathcal{C}^{\rm in})}{\partial \bar{y}} \right]^T.
	\end{align}
	
	Leveraging the chain rule, we have
	\begin{align}\label{eq:kan_gradient}
		& \frac{\partial \mathcal{F}_{\rm KAN}(\bar{\mathbf{q}}, \mathcal{C}^{\rm in})}{\partial \bar{x}} = \frac{\partial \mathcal{R}_{\rm KAN}(\mathbf{s}_{\rm reg})}{\partial \mathbf{s}_{\rm reg}} \cdot \frac{\partial \mathbf{s}_{\rm reg}}{\partial \bar{x}}, \\
		& \frac{\partial \mathcal{F}_{\rm KAN}(\bar{\mathbf{q}}, \mathcal{C}^{\rm in})}{\partial \bar{y}} = \frac{\partial \mathcal{R}_{\rm KAN}(\mathbf{s}_{\rm reg})}{\partial \mathbf{s}_{\rm reg}} \cdot \frac{\partial \mathbf{s}_{\rm reg}}{\partial \bar{y}},
	\end{align}
	where $\mathbf{s}_{\rm reg} = \bar{\mathbf{q}} \oplus \mathbf{s}$, and $\mathbf{s} = \mathcal{I}(\mathcal{C}^{\rm out}, 2\bar{\mathbf{q}} - 1)$. The Jacobian $\frac{\partial \mathbf{s}_{\rm reg}}{\partial \bar{x}}$ accounts for the dependency on both explicit coordinates and sampled features expressed as
	\begin{align}\label{eq:fusion_input_gradient}
		\frac{\partial \mathbf{s}_{\rm reg}}{\partial \bar{x}} = \begin{bmatrix} 1 \\ 0 \\ \frac{\partial \mathbf{s}}{\partial \bar{x}} \end{bmatrix}, \quad \frac{\partial \mathbf{s}_{\rm reg}}{\partial \bar{y}} = \begin{bmatrix} 0 \\ 1 \\ \frac{\partial \mathbf{s}}{\partial \bar{y}} \end{bmatrix},
	\end{align}
	since $\mathbf{s}_{\rm reg} = [\bar{x}, \bar{y}, \mathbf{s}^T]^T$.
	
	For the bilinear interpolation, with $\mu, \nu$ as fractional offsets in the grid cell $(i,j)$, and $\bar{\mathbf{q}}' = 2\bar{\mathbf{q}} - 1$. The partial derivatives can be calculated as
	\begin{align}\label{eq:gridding_gradient}
		\frac{\partial s_{(d)}}{\partial \bar{x}} = & 2(1-\nu)(\mathcal{C}^{\rm out}_{(d,i+1,j)} - \mathcal{C}^{\rm out}_{(d,i,j)}) \\ & \qquad \qquad \qquad + 2\nu(\mathcal{C}^{\rm out}_{(d,i+1,j+1)} - \mathcal{C}^{\rm out}_{(d,i,j+1)}), \notag
		\\
		\frac{\partial s_{(d)}}{\partial \bar{y}} = & 2 (1-\mu)(\mathcal{C}^{\rm out}_{(d,i,j+1)} - \mathcal{C}^{\rm out}_{(d,i,j)}) \\ & \qquad \qquad \qquad +
		2\mu(\mathcal{C}^{\rm out}_{(d,i+1,j+1)} - \mathcal{C}^{\rm out}_{(d,i+1,j)}), \notag
	\end{align}
	where the factor 2 arises from the normalization $\bar{\mathbf{q}}' = 2\bar{\mathbf{q}} - 1$. 
	
	The KAN regressor's gradient $\frac{\partial \mathcal{R}_{\rm KAN}(\mathbf{s}_{\rm reg})}{\partial \mathbf{s}_{\rm reg}}$ is computed by backpropagating through the spline layers. For each univariate function $\psi_{l,i,j}$, we have
	\begin{align}\label{eq:kan_B_spline_graidant}
		\frac{\partial \psi_{l,i,j}}{\partial v_i^{(l-1)}} = \sum_{m=0}^{g+k-2} c_{l,i,j,m} \frac{d B_{m,k}(v_i^{(l-1)})}{dv},
	\end{align}
	with
	\begin{align}\label{eq:kan_B_spline_graidant_extend}
		\frac{d B_{m,k}(v)}{dv} = (k-1) \left( \frac{B_{m,k-1}(v)}{t_{m+k-1} - t_m} - \frac{B_{m+1,k-1}(v)}{t_{m+k} - t_{m+1}} \right),
	\end{align}
	where terms with zero denominators are zero. Since B-splines are $C^{k-1}$-continuous, the KAN provides smooth gradients. The full gradient $\frac{\partial \mathcal{R}_{\rm KAN}}{\partial \mathbf{s}_{\rm reg}}$ is obtained via the chain rule across layers, leveraging the additive structure of KAN.
	
	This formulation ensures that $\nabla_{\bar{\mathbf{q}}} \hat{\mathcal{H}}(\bar{\mathbf{q}})$ is well-defined and computable, incorporating both explicit location dependencies. Unlike our conference work \cite{zhao2025diffkan}, which used only normalized coordinates, the inclusion of CNN-extracted features $\mathbf{s}$ enriches the gradient with site-specific information, enhancing its applicability to complex urban scenarios.
	This integrated encoder-regressor network forms a deterministic and fully differentiable computational graph. {Although the resultant overall gradient is piecewise-smooth, its jump discontinuities are confined to discrete grid boundaries forming a zero-measure set. In practice, automatic differentiation frameworks supply valid subgradients at these locations, ensuring that gradient-based solvers like SCA remain robust without compromising convergence.} Consequently, for any downstream task, the required gradient can be computed automatically and efficiently via backpropagation, treating the entire CKM construction model as a composable, gradient-aware function block.

	\section{Joint Power, Bandwidth, and Trajectory Optimization based on Differentiable CKM}
	This section studies trajectory design and power control for multiple UAVs by utilizing the cKAN-based differentiable CKMs in Section III. Specifically, we jointly optimize the UAV trajectories $\mathcal{Q}$, bandwidth allocation coefficients $\mathcal{A}$, and transmit powers $\mathcal{P}$ to maximize the minimum throughput among all UAVs over the flight period. We employ an AO algorithm combined with SCA to iteratively solve the problem.

	\vspace{-0.1in}
	\subsection{Power Optimization}
	When the UAV trajectories $\mathbf{q}_m[n] \in \mathcal{Q}$ and bandwidth allocation coefficients $\alpha_m[n] \in \mathcal{A}$ are fixed, the channel gains $h_m[n] = \mathcal{F}_{\rm KAN}(\mathbf{q}_m[n])$ become known constants. The optimization problem for transmit powers is formulated as
	\begin{align}\label{prob_power}
		{\rm (P1)}:	\quad \max_{{\mathcal{P}}} \quad & \min_m \left( \frac{1}{N} \sum_{n=1}^N R_m[n] \right) \\
		{\text{s.t.}} & \quad (\ref{prob0}\rm b), (\ref{prob0}\rm g) \notag
	\end{align}
	where the downlink transmission rate $R_m[n]$ for the $m$-th UAV at $n$-th time slot is given by (\ref{eq:approx_rate}), where $\mathcal{F}(\cdot)$ is replaced by cKAN network $\mathcal{F}_{\rm KAN}(\cdot)$.
	Let $K^P_m[n] = \frac{\mathcal{F}_{\rm KAN}(\mathbf{q}_m[n])}{N_0 \alpha_m[n] B_{\max}}$, we have
	\begin{align}\label{est_R_power}
		\hat{R}_m[n] & = \alpha_m[n] B_{\max} \log_2\left(1 + \frac{p_m[n] \mathcal{F}_{\rm KAN}(\mathbf{q}_m[n])}{N_0 \alpha_m[n] B_{\max}}\right) \\
		& = \alpha_m[n] B_{\max} \log_2(1 + K^P_m \cdot p_m[n]) = \hat{\mathcal{H}}(\mathbf{q}_m[n]). \notag
	\end{align}

	Since (\ref{est_R_power}) is a concave function with respect to $p_m[n]$, and $A, C > 0$, $R_m[n]$ is a concave function of $p_m[n]$. The objective function is the minimum of a sum of concave functions; hence, it remains concave.


	\vspace{-0.1in}
	\subsection{Bandwidth Optimization}
	With fixed UAV trajectories $\mathbf{q}_m[n] \in \mathcal{Q}$ and transmit powers $\mathbf{p}_m[n] \in \mathcal{P}$, we optimize the bandwidth allocation coefficients $\alpha_m[n] \in \mathcal{A}$. For each time slot $n$, the optimization problem is expressed as
	\begin{align}\label{prob_bandwidth}
		({\rm P2}):\quad \max_{\mathcal{A}} \quad & \min_m \left( \frac{1}{N} \sum_{n=1}^N R_m[n] \right) \\
		{\rm s.t.} \quad & R_m[n] = \hat{R}_m[n], \tag{\ref{prob_bandwidth}a}\\
		& (\ref{prob0}\rm a), (\ref{prob0}\rm g) \notag
	\end{align}
	Let $K^B_m[n] = \frac{p_m[n] \mathcal{F}_{\rm KAN}(\mathbf{q}_m[n])}{N_0 B_{\max}}$, we have
	\begin{align}\label{est_R_bandwidth}
		\hat{R}_m[n] = \hat{\mathcal{H}}(\mathbf{q}_m[n])= \alpha_m[n] B_{\max} \log_2\left(1 + \frac{K^B_m[n]}{\alpha_m[n]}\right).
	\end{align}
	The second derivative of $R_m[n]$ with respect to $\alpha_m[n]$ is
	\begin{align}
		\frac{d^2 \hat{R}_m[n]}{d\alpha_{m}[n]^2} = -\frac{B_{\max} K^B_m[n]^2}{\alpha_m[n](K^B_m[n]+\alpha_m[n])^2 \ln{2} } < 0,
	\end{align}
	which confirms that $\hat{R}_m[n]$ in (\ref{est_R_bandwidth}) is a concave function of $\alpha_m[n]$. However, due to the quotient form $K^B_m[n]/\alpha_m[n]$ within the logarithm, this expression is not in a directly convex-tractable form.
	To address this, we employ the SCA method. In the $l$-th iteration of the SCA process, we perform a first-order Taylor expansion of $R_m[n]$ around the current bandwidth allocation coefficients $\alpha_m^{(l)}[n]$ to obtain its lower bound, the lower bound approximation $\breve{R}_m[n]$ for $R_m[n]$ is given by:
	\begin{align}\label{eq:SCA_band}
		\hat{R}_m[n] \geq \hat{R}_m^{(l)}[n] + \nabla_{\mathbf{\alpha}} \hat{R}^{(l)}_m[n] (\alpha_m[n] - \alpha_m^{(l)}[n]) \triangleq \breve{R}_m[n].
	\end{align}

	Here, $R_m^{(l)}[n]$ represents the value of $R_m[n]$ calculated using $\alpha_m^{(l)}[n]$.
	Let $\nabla_{\mathbf{\alpha}} \hat{R}_m[n]$ be the first derivative of $\hat{R}_m[n]$ with respect to $\alpha_m[n]$, evaluated at $\alpha_m^{(l)}[n]$, as expressed as
	\begin{align}
		& \nabla_{\mathbf{\alpha}} \hat{R}_m[n] = \frac{\partial \hat{R}_m[n]}{\partial \alpha_m[n]} \bigg|_{\alpha_m[n]=\alpha_m^{(l)}[n]} \\
		& \qquad \,\, = B \left[ \log_2\left(1 + \frac{K^B_m[n]}{\alpha_m^{(l)}[n]}\right) - \frac{K^B_m[n]}{(\alpha_m^{(l)}[n] + K^B_m[n])\ln 2} \right]. \notag
	\end{align}
	By introducing an auxiliary variable $\gamma_b$, the original problem is transformed into a convex formula as expressed as
	\begin{align}\label{prob_band_2}
		({\rm P2}'): \quad \max_{\mathcal{A}} \quad & \gamma_b \\
		\text{s.t.} \quad & \frac{1}{N} \sum_{n=1}^N \breve{R}_m[n] \geq \gamma_b, \quad \forall m \tag{\ref{prob_band_2}a} \\
		& \breve{R}_m[n] \geq R_{\text{min}}, \quad \forall m, n \tag{\ref{prob_band_2}b} \\
		& (\ref{prob0}\rm a) \notag
	\end{align}
	This reformulated problem $({\rm P2}')$ is convex and can be efficiently solved in each SCA iteration.

	\vspace{-0.1in}
	\subsection{Trajectory Optimization}
	With transmit powers $p_m[n] \in \mathcal{P}$ and bandwidth allocation coefficients $\alpha_m[n] \in \mathcal{A}$ fixed, we optimize the UAV trajectories $\mathbf{q}_m[n] \in \mathcal{Q}$. The problem is formulated as
	\begin{align}\label{prob_traj}
		({\rm P3}): \quad \max_{\mathcal{Q}} \quad & \min_m \left( \frac{1}{N} \sum_{n=1}^N R_m[n] \right) \\
		{\rm s.t.} \quad & R_m[n] = \hat{R}_m[n] \tag{\ref{prob_traj}a} \\
		& \Vert\mathbf{q}_m[n] - \mathbf{q}_{{\rm obs},k}\Vert_2 \geq D_{\min}, \,\, \forall m, n, k \tag{\ref{prob_traj}b} \\
		& \mathbf{q}_m[n] \in \mathcal{D}, \,\, \forall m, n \tag{\ref{prob_traj}c} \\
		& (\ref{prob0}\rm c), (\ref{prob0}\rm d), (\ref{prob0}\rm g) \notag
	\end{align}
	Both the rate constraint (\ref{prob_traj}a) and the obstacle avoidance constraint (\ref{prob_traj}b) are non-convex. We employ the SCA method to transform these non-convex constraints into a series of convex ones. The specific approximation is shown as follows.

	\subsubsection{Approximation of constraint (\ref{prob_traj}a)}
	Initial trajectories can be set as straight lines. In the $l$-th iteration, based on the current coordinate of trajectory points $\mathbf{q}_m^{(l)}[n]$, we then perform a first-order Taylor expansion of $R_m[n]$ to obtain its lower bound as expressed as
	\begin{align}\label{eq:SCA_traj}
		\hat{R}_m[n] \geq \hat{R}_m^{(l)}[n] + \nabla_{\mathbf{q}} \hat{R}_m[n]^T (\mathbf{q}_m[n] - \mathbf{q}_m^{(l)}[n]) \triangleq \breve{R}_m[n],
	\end{align}
	where $\nabla_{\mathbf{q}} \hat{R}_m[n]$ is the gradient of the rate $\hat{R}_m[n]$ with respect to the trajectory $\mathbf{q}_m[n]$. This gradient is found via the chain rule as introduced in Section III. C as
	\begin{align}
		\nabla_{\mathbf{q}} \hat{R}_m[n] = \frac{\partial \hat{R}_m[n] \cdot \nabla_{\mathbf{q}} \mathcal{F}_{\rm KAN}(\mathbf{q}_m[n])}{\partial \mathcal{F}_{\rm KAN}(\mathbf{q}_m[n])} \bigg|_{\mathbf{q}_m[n]=\mathbf{q}_m^{(l)}[n]}.
	\end{align}
	With $E^{\mathcal{F}}_m = \frac{\partial R_m[n]}{\partial \mathcal{F}_{\rm KAN}(\mathbf{q}_m[n])}$, we have
	\begin{align}\label{eq:gradient_R_q}
		E^{\mathcal{F}}_m = \frac{\alpha_m[n] B_{\max}}{\ln 2} \cdot \frac{p_m[n]}{N_0 \alpha_m[n] B_{\max} + p_m[n] \mathcal{F}_{\rm KAN}(\mathbf{q}_m[n])}.
	\end{align}

	The gradient of the cKAN, i.e., $\nabla_{\mathbf{q}} \mathcal{F}_{\rm KAN}(\mathbf{q}_m[n])$, is crucial. With $\mathbf{q}_m[n] = [x_m[n], y_m[n]]^T$ being the normalized 2D position coordinates, the gradient can be expressed as
	\begin{align}\label{eq:gradient_F_q}
		\nabla_{\mathbf{q}} \mathcal{F}_{\rm KAN}(\mathbf{q}_m[n]) = \left[ \frac{\partial \mathcal{F}_{\rm KAN}}{\partial x_m[n]}, \frac{\partial \mathcal{F}_{\rm KAN}}{\partial y_m[n]} \right]^T,
	\end{align}
	where the partial derivatives are obtained by applying the chain rule through the entire cKAN architecture, which are shown in (\ref{eq:kan_gradient}).
	With $\mathbf{s}_{\rm reg} = [x_m[n], y_m[n], \mathbf{s}^T]^T$. The partial derivatives of $\mathbf{s}_{\rm reg}$ are shown in (\ref{eq:fusion_input_gradient}). The components $\frac{\partial \mathbf{s}}{\partial x_m[n]}$ and $\frac{\partial \mathbf{s}}{\partial y_m[n]}$ are derived from the bilinear interpolation function $\mathcal{I}(\cdot)$ as provided in (\ref{eq:gridding_gradient}), while $\frac{\partial \mathcal{R}_{\rm KAN}}{\partial \mathbf{s}_{\rm reg}}$ is the gradient of the KAN regressor, computed via backpropagation through its spline layers, utilizing the derivatives of B-spline basis functions also described in (\ref{eq:kan_B_spline_graidant}) and (\ref{eq:kan_B_spline_graidant_extend}).

	It is worth noting that the proposed CKM-JPBTO framework is model-agnostic regarding the regression backbone. While this paper focuses on the cKAN-based regressor due to its superior fitting accuracy and lightweight deployment, other differentiable architectures, such as the cMLP / MLP baseline, can also be integrated. For cMLP, the gradient $\nabla_{\mathbf{q}} \mathcal{F}$ can be similarly computed via standard backpropagation as derived in \cite{lyMLPCKM2025tvt}, allowing it to serve as a valid alternative for trajectory optimization.

	Now, we have the gradient of $\hat{R}_m[n]$ with respect to $\mathbf{q}$ based on (\ref{eq:gradient_R_q}) and (\ref{eq:gradient_F_q}), which can be expressed as
	\begin{align}
		\nabla_{\mathbf{q}} \hat{R}_m[n] = E^{\mathcal{F}}_m \cdot  \nabla_{\mathbf{q}} \mathcal{F}_{\rm KAN}(\mathbf{q}_m[n]),
	\end{align}
	and the approximation of constraint (\ref{prob_traj}a) is obtained.

	\subsubsection{Approximation of constraint (\ref{prob_traj}b) and (\ref{prob_traj}c)}
	By squaring both sides of constraint (\ref{prob_traj}b), we have
	\begin{align}
		|\mathbf{q}_m[n] - \mathbf{q}_{{\rm obs},k}|_2^2 \geq (D_{\min})^2.
	\end{align}
	Applying a first-order Taylor expansion around $\mathbf{q}_m^{(l)}[n]$ to the left-hand side, we obtain the following convex approximation:
	\begin{align}\label{eq:SCA_obs}
		& |\mathbf{q}_m[n] - \mathbf{q}_{{\rm obs},k}|_2^2 \geq |\mathbf{q}_m^{(l)}[n] - \mathbf{q}_{{\rm obs},k}|_2^2  \\
		& \qquad \qquad + 2 (\mathbf{q}_m^{(l)}[n] - \mathbf{q}_{{\rm obs},k})^T (\mathbf{q}_m[n] - \mathbf{q}_m^{(l)}[n])
		\triangleq \breve{D}_m[n]. \notag
	\end{align}

	We then transform the constraint (\ref{prob_traj}c) to linear formulation, with $\mathbf{q}_{\min} = [X_{\min}, Y_{\min}]^T$, and $\mathbf{q}_{\max} = [X_{\max}, Y_{\max}]^T$. The trajectory optimization problem is transformed into
	\begin{align}\label{prob_traj_SCA}
		({\rm P3}'): \quad \max_{\mathcal{Q}} \quad & \gamma_q \\
		{\rm s.t.} \quad & \frac{1}{N} \sum_{n=1}^N \breve{R}_m[n] \geq \gamma_q, \quad \forall m \tag{\ref{prob_traj_SCA}a}\\
		& \breve{R}_m[n] \geq R_{\text{min}}, \quad \forall m, n \tag{\ref{prob_traj_SCA}b}\\
		& \mathbf{q}_{\text{min}} \leq \mathbf{q}_m[n] \leq \mathbf{q}_{\text{max}}, \quad \forall m, n \tag{\ref{prob_traj_SCA}c}\\
		& \breve{d}_m[n] \geq D_{\min} \quad \forall m, n \tag{\ref{prob_traj_SCA}d} \\
		& (\ref{prob0}\rm c), (\ref{prob0}\rm d) \notag
	\end{align}
	This reformulated problem is convex and can be solved in each SCA iteration.

	\begin{algorithm}[!t]
		\caption{Proposed CKM-JPBTO algorithm}
		\label{alg:overall_algo}
		\SetAlgoLined
		\textbf{Input}: {UAV start positions $\mathbf{q}_m^{\text{start}}$ and end positions $\mathbf{q}_m^{\text{end}}$, maximum power $P_{\text{max}}$, maximum speed $V_{\text{max}}$, noise power spectral density $N_0$, time slot duration $\delta$.}\\
		\textbf{Initialize}: iteration count $l=0$, set maximum AO iterations $L_{\rm max}$, AO convergence threshold $\epsilon_{\rm AO}$, trajectories $\mathcal{Q}^{(0)}$ as linear interpolation, bandwidth $\mathcal{A}^{(0)}$ as $\alpha_m[n] = 1/M$, transmit powers $\mathcal{P}^{(0)}$ as $p_m[n] = P_{\max}$. \\
		\Repeat{$l = L_{\max}$ or $\left\Vert \mathcal{Q}^{(l, i)} - \mathcal{Q}^{(l, i-1)} \right\Vert \leq \epsilon_{\rm AO}$}{
			\textit{1) -- -- Power optimization -- --} \\
			Solve the problem $({\rm P1})$ in (\ref{prob_power}) with fixed $\mathcal{Q}^{(l)}$ and $\mathcal{A}^{(l)}$, obtain optimal powers $\{p_m^{(l+1)}[n]\} \in \mathcal{P}^{(l+1)}$.\\
			\textit{2) -- -- Bandwidth optimization -- --} \\
			\textbf{Initialize}: iteration count $i=0$, maximum iterations $I_{\rm max}$, convergence threshold $\epsilon_{\alpha}$.\\
			\Repeat{$i = I_{\max}$ or $\left\Vert \mathcal{A}^{(l, i)} - \mathcal{A}^{(l, i-1)} \right\Vert \leq \epsilon_{\alpha}$}{
				Perform SCA, obtain $\breve{R}_m[n]$ based on (\ref{eq:SCA_band}).\\
				Solve the problem $({\rm P2}')$ in (\ref{prob_band_2}) with fixed $\mathcal{Q}^{(l)}$ and $\mathcal{P}^{(l+1)}$, obtain optimal $\{\alpha_m[n]^{(l, i+1)}\} \in \mathcal{A}^{(l,i+1)}$.\\
				Update $i \leftarrow i+1$.
			}
			Set $\mathcal{A}^{(l+1)} = \mathcal{A}^{(l,i)}$.\\
			\textit{3) -- -- Trajectory Optimization -- --}\\
			\textbf{Initialize}: iteration count $j=0$, set maximum iterations $J_{\rm max}$, convergence threshold $\epsilon_{q}$.\\
			\Repeat{$j = J_{\max}$ or $\left\Vert \mathcal{Q}^{(l, i)} - \mathcal{Q}^{(l, i-1)} \right\Vert \leq \epsilon_{q}$}{
				Perform SCA, obtain $\breve{R}_m[n]$ by (\ref{eq:SCA_traj}) and obstacle avoidance constraints by $\mathcal{Q}^{(l,j)}$ based on (\ref{eq:SCA_obs}).\\
				Solve the problem $({\rm P3}')$ in (\ref{prob_traj_SCA}) with fixed $\mathcal{A}^{(l+1)}$ and $\mathcal{P}^{(l+1)}$, obtain optimal $\{\mathbf{q}_m[n]^{l, j+1}\} \in \mathcal{Q}^{(l,j+1)}$.\\
				Update $j \leftarrow j+1$.
			}
			Set $\mathcal{Q}^{(l+1)} = \mathcal{Q}^{(l,j)}$.\\
			Update $l \leftarrow l+1$.
		}
		\textbf{Output}: Optimal $\mathcal{Q}^* = \mathcal{Q}^{(l)}$, $\mathcal{A}^* = \mathcal{A}^{(l)}$, $\mathcal{P}^* = \mathcal{P}^{(l)}$.
	\end{algorithm}
	
	\subsection{Overall Algorithm}
	The overall algorithm for problem (P0) iteratively solves subproblems (P1)–(P3) in an alternating fashion, with the output of each iteration serving as the input for the next. As summarized in \myrefalgo{alg:overall_algo}, convergence is guaranteed by two key properties: 1) the successive convex approximations via Taylor expansion yield a non-decreasing objective sequence, and 2) the objective is upper-bounded due to physical constraints on UAV speed and transmit power. The algorithm converges to a stationary point by solving convex subproblems for power$/$bandwidth and applying SCA to the trajectory subproblem, with termination triggered by minimal objective improvement or reaching the maximum iterations.

	The complexity analysis focuses on the regressor inference and the iterative optimization. Since the CNN-based feature extraction is performed once and cached, the recurring cost depends on the regressor.
	The cMLP baseline scales as $\mathcal{O}(L_{\rm M} W_{\rm M}^2)$, where $L_{\rm M}$ and $W_{\rm M}$ denote depth and width.
	In contrast, the proposed cKAN scales as $\mathcal{O}(L_{\rm K} W_{\rm K}^2 (g+k))$, introducing a linear dependency on grid size $g$ and spline order $k$. The CKM-JPBTO algorithm employs an AO framework with $L_{\max}$ outer iterations. In each iteration, the power, bandwidth, and trajectory subproblems are solved sequentially. The trajectory optimization is the computational bottleneck, as it involves solving a series of convex programming problems via SCA ($J_{\max}$ iterations) as well as deriving gradients via backpropagation through the regressor for all $M$ UAVs over $N$ time slots. The complexity analysis of the proposed algorithm is summarized in \myreftable{tab:complexity}, where $C_{\rm Reg}$ denotes the per-sample computational cost of a single forward and backward pass through the chosen regressor.

	\begin{table}[!t]
		\caption{Complexity Comparison of Different Components}
		\label{tab:complexity}
		\centering
		\setlength{\tabcolsep}{2mm}
		\renewcommand{\arraystretch}{1.15}
		\begin{tabular}{l | l}
			\hline
			\textbf{Component} & \textbf{Computational Complexity} \\
			\hline
			CNN encoder & $\mathcal{O}(L_{\rm CNN} \cdot (d^{\rm out})^2 \cdot H^{\rm in} W^{\rm in} \cdot K^2)$ \\
			cMLP regressor & $\mathcal{O}(L_{\rm MLP} \cdot W_{\rm MLP}^2)$ \\
			cKAN regressor & $\mathcal{O}(L_{\rm KAN} \cdot W_{\rm KAN}^2 \cdot (g + k))$ \\
			\hline
			\textbf{Overall Algorithm} & $\mathcal{O}\Big( L_{\max} \big[ (I_{\max} {+} J_{\max})(MN)^{3.5}$ \\
			& $\, + \, J_{\max} MN C_{\rm reg} \big] \Big)$ \\
			\hline
		\end{tabular}
	\end{table}
	
	\section{Numerical Results}
	In this section, we firstly present numerical results to calibrate the performance of the proposed cKAN-based differentiable CKM construction. Subsequently, we provide simulation results to demonstrate the effectiveness of the proposed CKM-JPBTO algorithm for multi-UAV systems.

	\vspace{-0.15in}
	\subsection{Simulation Setup}
	The training dataset is generated using Wireless InSite \cite{WirelessInSite}, a ray-tracing tool. We consider a $\mathcal{D}=2000 \times 2000$ square meters area in the central Beijing\footnote{The 3D city map is available at http://www.openstreetmap.org.}. The BS operates at a height of 25 m, serving UAVs at $z_h = 100$ m. The building distribution is shown in Fig. 3.
	The simulation operates at 2.4 GHz, modeling path loss, shadowing, with reflection and diffraction of buildings. The original CKM contains $X \times Y=256\times256$ unit areas, each with the size of $\Delta \kappa = 7.8125$ square meters, which is chosen to be sufficiently small such that the channel gain within each grid is approximately constant. We collect $256^2$ channel gain measurements of discrete locations set $\tilde{\mathcal{D}}$ to evaluate the CKM construction error and communication performance of a multi-UAV system. We randomly sampled the training subset from $\tilde{\mathcal{D}}$ with ratios $[0.5\%, 3\%]$, forming $\{ \mathcal{C}, \mathbf{q}, \mathcal{H}(\mathbf{q}) \}, \mathbf{q} \in \mathcal{D}^s$ for model training, and the remaining subset is used for model evaluation. The networks are trained using the Adam optimizer with a learning rate of $10^{-3}$ and a batch size of 128 over 200 epochs. During the JPBTO procedure, the convergence threshold $\epsilon_{\rm q}$, $\epsilon_{\rm \alpha}$, $\epsilon_{\rm AO}$ are set as $10^{-4}$, $\epsilon$ in \myrefeq{prob0} is set as $10^{-6}$, and the maximum iterations are set as $I_{\max} = J_{\max} = 50, L_{\max} = 15$.

	\begin{figure}[!t]
		\centering
		\includegraphics[width=0.7\columnwidth]{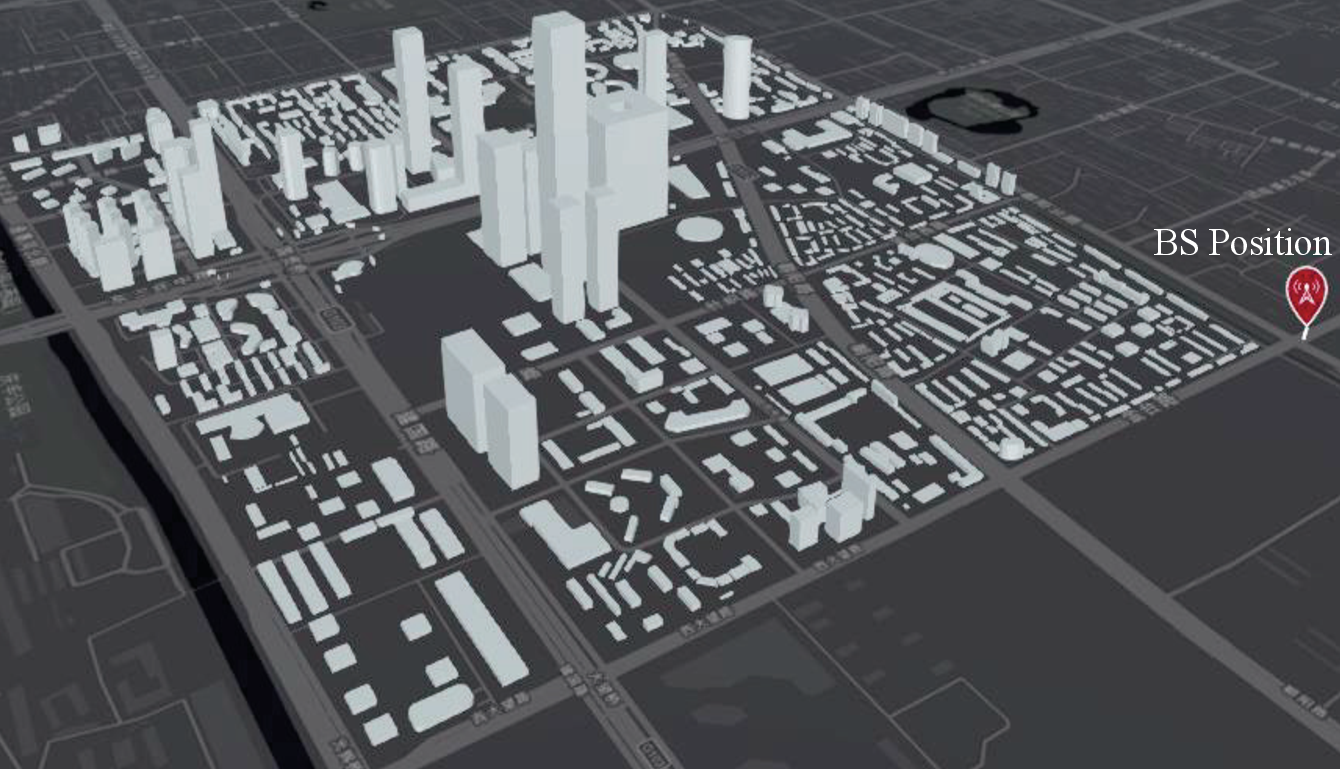}
		\caption{Visualization of building distribution in simulation.}
		\label{fig:simulation_area}
		\vspace{-0.1in}
	\end{figure}

	\begin{table}[!t]
		\renewcommand{\arraystretch}{1.05}
		\setlength{\tabcolsep}{1.5mm}
		\centering
		\caption{Architectural of CKM construction networks}
		\label{tab:model_configs}
		\begin{tabular}{l|l}
			\hline
			\textbf{Model} & \textbf{Core Architecture} \\
			\hline
			\textbf{MLP}\,/\,\textbf{Ka-MLP} & \texttt{fc1}: Linear(2, 64)\\
			& \texttt{fc2}: Linear(64, 128)\\
			& \texttt{fc3}: Linear(128, 64)\\
			& \texttt{fc4}: Linear(64, 32)\\
			& \texttt{fc5}: Linear(32, 1) \\
			\hline
			\textbf{KAN}\,/\,\textbf{Ka-KAN} & KAN(width=[2, 10, 20, 10, 1], grid=10, $k=4$) \\
			\hline
			\textbf{cMLP} & \textbf{CNN Encoder:} \\
			& \quad \texttt{conv1}: Conv2d(4, 16, $5\times5$, padding=2)\\
			& \quad \texttt{res1}: ResidualBlock(16, 32) $\rightarrow$ MaxPool2d(2) \\
			& \quad \texttt{conv2}: Conv2d(32, 64, $3\times3$, padding=2)\\
			& \quad \texttt{res2}: ResidualBlock(64, 128) $\rightarrow$ MaxPool2d(2) \\
			& \quad \texttt{conv3}: Conv2d(128, 128, $3\times3$, padding=1)\\
			& \quad \texttt{res3}: ResidualBlock(128, 128) $\rightarrow$ MaxPool2d(2) \\
			& \textbf{MLP Regressor:} \\
			& \quad \texttt{fc1}: Linear(130, 128)\\
			& \quad \texttt{fc2}: Linear(128, 32)\\
			& \quad \texttt{fc3}: Linear(32, 1) \\
			\hline
			\textbf{cKAN} & \textbf{CNN Encoder:} \\
			& \quad \texttt{conv1}: Conv2d(4, 16, $5\times5$, padding=2)\\
			& \quad \texttt{res1}: ResidualBlock(16, 32) $\rightarrow$ MaxPool2d(2) \\
			& \quad \texttt{conv2}: Conv2d(32, 64, $3\times3$, padding=2)\\
			& \quad \texttt{res2}: ResidualBlock(64, 128) $\rightarrow$ MaxPool2d(2) \\
			& \quad \texttt{conv3}: Conv2d(128, 64, $3\times3$, padding=1)\\
			& \quad \texttt{res3}: ResidualBlock(64, 64) $\rightarrow$ MaxPool2d(2) \\
			& \textbf{KAN Regressor:} \\
			& \quad KAN(width=[66, 10, 1], grid=8, $k=4$) \\
			\hline
		\end{tabular}
		\vspace{-0.15in}
	\end{table}

	\begin{figure*}[!t]
		\centering
		\subfloat[]{\includegraphics[width=1.5in]{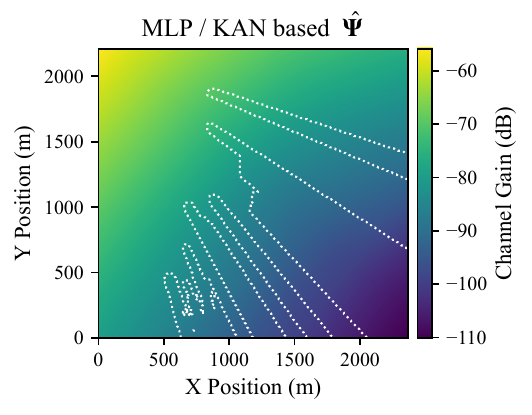}}
		\subfloat[]{\includegraphics[width=1.5in]{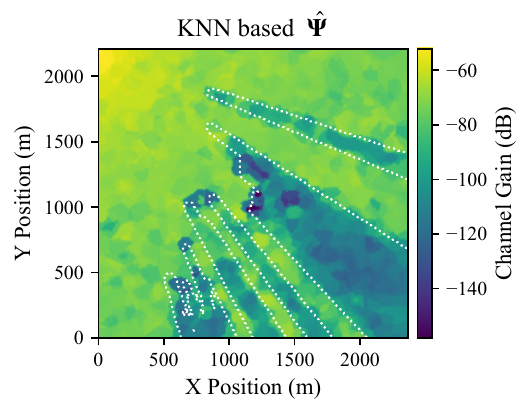}}
		\subfloat[]{\includegraphics[width=1.5in]{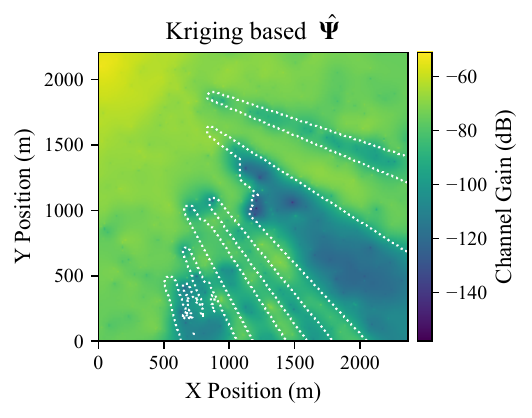}}
		\subfloat[]{\includegraphics[width=1.5in]{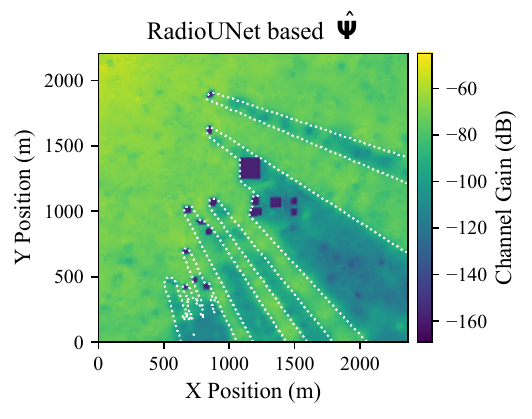}}
		\vspace{-0.15in}  
		\hspace{0.002in}
		\subfloat[]{\includegraphics[width=1.5in]{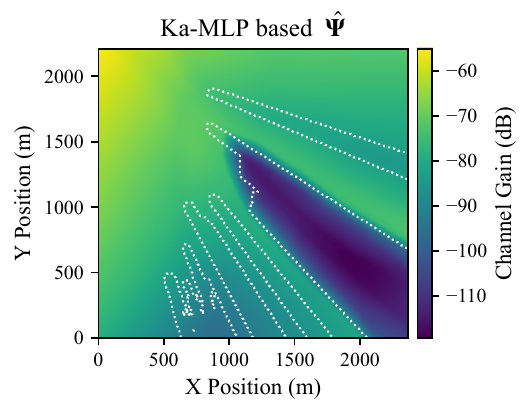}}
		\subfloat[]{\includegraphics[width=1.5in]{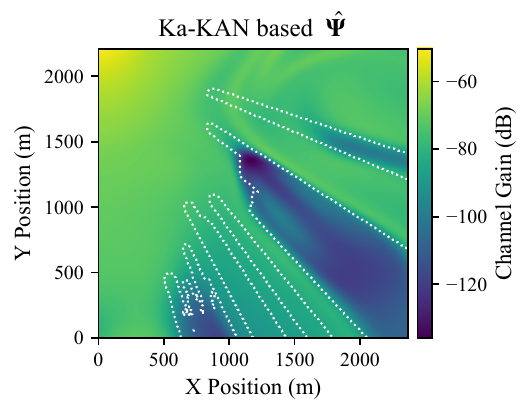}}
		\subfloat[]{\includegraphics[width=1.5in]{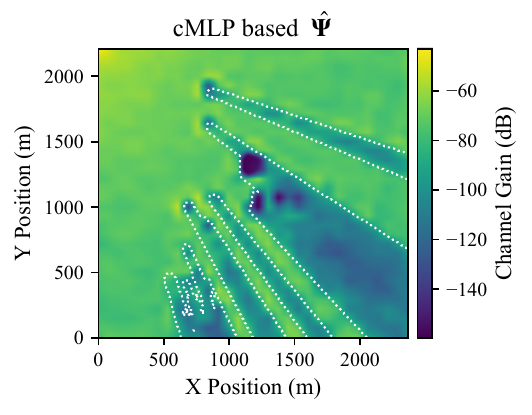}}
		\subfloat[]{\includegraphics[width=1.5in]{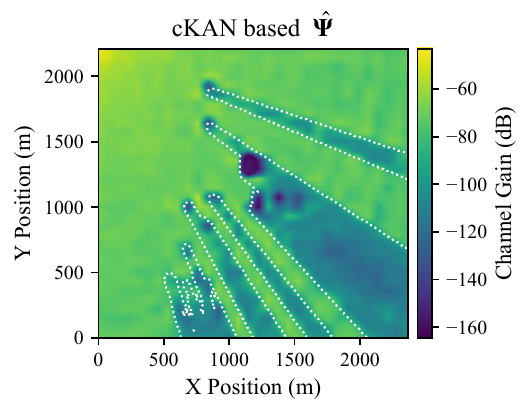}}
		\caption{Visual comparison of CKM predictions using: (a) MLP \cite{lyMLPCKM2025tvt} / KAN \cite{zhao2025diffkan}, (b) KNN \cite{KNN_2009_TIP}, (c) Kriging \cite{Krige_2017_TVT}, (d) RadioUNet \cite{Levie_2021_TWC}, (e) Ka-MLP \cite{zhao2025diffkan}, (f) Ka-KAN \cite{zhao2025diffkan}, (g) cMLP, and (h) cKAN. The sampling ratio is set to ${\left\vert \mathcal{D}^s \right\vert}/{\vert \mathcal{\tilde{D}} \vert} = 3\%$, with the LoS/NLoS boundary indicated by the white dashed line. The ground-truth $\mathbf{\Psi}$ is provided in \myreffig{fig:trajectories_CKM}-\myreffig{fig:trajectories_LoS}.}
		\label{fig:predicted_CKM_visualization}
		\vspace{-0.2in}
	\end{figure*}
	
	\vspace{-0.1in}
	\subsection{CKM construction}
	To demonstrate the CKM construction performance, we construct the predicted CKM $\hat{\mathbf{\Psi}}$ with the set of discrete locations $\tilde{\mathcal{D}}$.
	To comprehensively evaluate the quality of the constructed CKM, we adopt the normalized mean square error (NMSE) commonly used in previous studies, defined as
	\begin{align}
		\label{nmse_define}
		{\rm NMSE} = \frac{\sum_{x=1}^{N_1}\sum_{y=1}^{N_2}\left( \widehat{\Psi}_{(x,y)}  - \Psi_{(x,y)} \right)^2}{\sum_{x=1}^{N_1}\sum_{y=1}^{N_2}\left( \Psi_{(x,y)} \right)^2}.
	\end{align}

	We compare our proposed differentiable CKM construction under different channel gain measurement sampling ratios with the following baselines:
	\begin{enumerate}
		\item[-] {KNN \cite{KNN_2009_TIP}}: Calculate the channel gain $\hat{\mathcal{H}}(\mathbf{q})$ based on the mean of the measured channel gain of adjacent locations in $\mathcal{D}^s$. The number of adjacent channel gain samples to calculate $\hat{\mathcal{H}}_(\mathbf{q})$ can be set as any positive integer. Note that predicted CKM by KNN is not differentiable with respect to $\mathbf{q}$.
		\item[-] {Kriging (linear kernel) \cite{Krige_2017_TVT}}: Calculate the channel gain $\hat{\mathcal{H}}(\mathbf{q})$ based on the Kriging-based spatial interpolation algorithm with linear kernel to maintain the differentiability with respect to $\mathbf{q}$. The predicted CKM  is location-differentiable.
		\item[-] {MLP \cite{lyMLPCKM2025tvt}}: Construct the CKM by an MLP network trained by the channel gain samples of locations $\mathbf{q} \in \mathcal{D}^s$, and is location-differentiable.
		\item[-] {KAN \cite{zhao2025diffkan}}: Replaces the MLP with KAN to predict channel gain, obtains a locatoin-differentiable CKM.
		\item[-] {Ka-MLP \cite{zhao2025diffkan}}: Construct the CKM by MLP network trained by the channel gain samples of locations $\mathbf{q} \in \mathcal{D}^s$ and the predicted channel gain $\hat{\mathcal{H}}(\mathbf{q}) \in \mathcal{D}-\mathcal{D}^s$ by KNN, and is location-differentiable CKM.
		\item[-] {Ka-KAN \cite{zhao2025diffkan}}: Construct the CKM by KAN network trained by the same training dataset as Ka-MLP, obtains a location-aware differentiable CKM.
		\item[-] RadioUNet \cite{Levie_2021_TWC}: Construct the CKM by CNN-based architecture U-Net, takes the environment topology, the sampled sparse CKM $\mathbf{\Psi}^{\rm s}$, and the BS location map as inputs, and outputs the predicted grid-based CKM $\hat{\mathbf{\Psi}}$.
	\end{enumerate}
	Configureation of the CKM construction methods is presented in \myreftable{tab:model_configs}.


	\myreffig{fig:predicted_CKM_visualization} presents a visual comparison of CKMs reconstructed by different methods under sampling ratio $\vert\mathcal{D}\vert / \vert\tilde{\mathcal{D}}\vert = 3\%$. Each subfigure displays a $256\times256$ grid heatmap, where color intensity represents the predicted channel gain, ranging from low (dark blue) to high (yellow). It is observed that the results of MLP\,/\,KAN-based CKM construction results in \myreffig{fig:predicted_CKM_visualization} (a) cannot work under such low channel sampling ratio. The KNN argumented Ka-MLP\,/\, Ka-KAN in \myreffig{fig:predicted_CKM_visualization} (e), (f), on the other hand, only learn the general distribution of channel gain, which is similar to KNN and Kriging in \myreffig{fig:predicted_CKM_visualization} (b), (c). In \myreffig{fig:predicted_CKM_visualization} (d), it can be observed that RadioUNet achieves the fine visual reconstruction, particularly in capturing the shadow boundaries behind buildings. Fortunately, with CNN-based feature extraction, both cMLP and cKAN achieves astonishing performance. As shown in \myreffig{fig:predicted_CKM_visualization} (g), (h), they clearly distinguished the LoS and NLoS regions, and also gained a limited understanding of the wave and reflection propagation.

	\begin{figure}[!t]
		\centering
		\includegraphics[width=0.73\columnwidth]{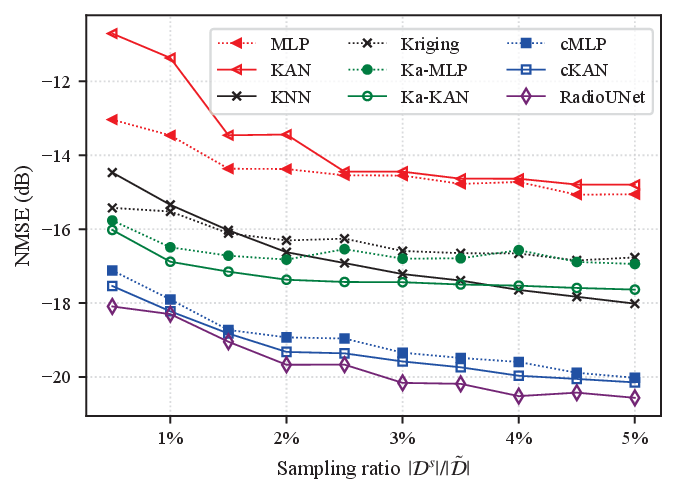} 
		\caption{NMSE of reconstructed CKM with different channel gain measurements sampling ratio ${\left\vert \mathcal{D}^s \right\vert}/{\vert \mathcal{\tilde{D}} \vert}$.}
		\label{fig:NMSE_plot}
		\vspace{-0.1in}
	\end{figure}
	
	\begin{figure}[!t]
		\centering
		\includegraphics[width=0.725\columnwidth]{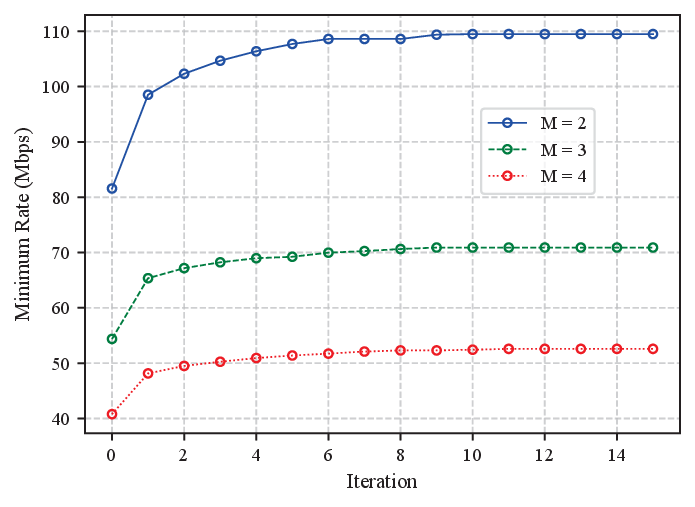}
		\caption{Convergence of CKM-JPBTO with $P_{\rm max}=10$ W, $B_{\max} = 10$ MHz, , $N = 50$, and $T=100$ s, by ${\left\vert \mathcal{D}^s \right\vert}/{\vert \mathcal{\tilde{D}} \vert} = 2\%$ sampled cKAN.}
		\label{fig:Convergence}
		\vspace{-0.1in}
	\end{figure}

	\begin{table}[!t]
		\caption{Empirical Latency Comparison}
		\label{tab:runtime}
		\centering
		\setlength{\tabcolsep}{3mm}
		\renewcommand{\arraystretch}{1.05}
		\begin{tabular}{l|l|c}
			\hline
			\textbf{Operation Stage} & \textbf{Method} & \textbf{Latency (s)} \\
			\hline
			{CKM Inference} & RadioUNet \cite{Levie_2021_TWC} & 0.058 \\
			& {cKAN (Ours)} & \textbf{0.035} \\
			\hline
			Gradient Computation & Backpropagation & 0.028 \\
			\hline
			{Optimization (Per Iteration)} & GWO-JPBTO & 0.888 \\
			& CKM-JPBTO & \textbf{0.293} \\
			\hline
		\end{tabular}
		\vspace{-0.1in}
	\end{table}
	
	\myreffig{fig:NMSE_plot} illustrates the NMSE performance of the discussed CKM construction methods. It is observed that the classic RadioUNet achieves the lowest NMSE due to its powerful U-Net architecture. Conversely, purely coordinate-based methods, including KNN, Kriging, and standard MLP or KAN, suffer from limited accuracy as they fail to capture complex environmental spatial relationships. In contrast, our proposed cMLP and cKAN methods effectively bridge this gap. By integrating environmental features, they achieve accuracy levels highly comparable to RadioUNet, and significantly superior to KNN and Kriging. Most importantly, when comparing the two proposed methods, cKAN achieves slightly better performance than cMLP with a significantly more lightweight structure. As shown in \myreftable{tab:model_configs}, the KAN regressor requires fewer learnable parameters than the MLP-based methods to model the complex channel mapping.
	
	{Beyond reconstruction accuracy, practical trajectory optimization demands efficient and reliable gradients. Although applying bilinear interpolation to the discrete output of RadioUNet could theoretically provide gradients, its heavy decoder necessitates full-grid reconstruction even for local coordinate queries. This incurs unnecessary computational latency, as verified in \myreftable{tab:runtime}. Conversely, while Kriging is inherently differentiable, its high NMSE produces unreliable gradients that misguide optimization. By providing an accurate, low-latency, and directly differentiable mapping, cKAN overcomes these bottlenecks, seamlessly enabling the trajectory optimization evaluated next.}
	
	\begin{figure}[!t]
		\centering
		\subfloat[]{\includegraphics[width=0.33\columnwidth]{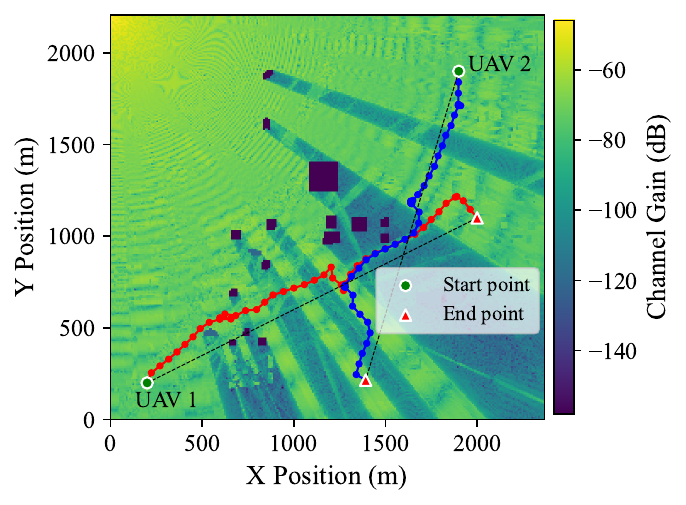}}
		\subfloat[]{\includegraphics[width=0.33\columnwidth]{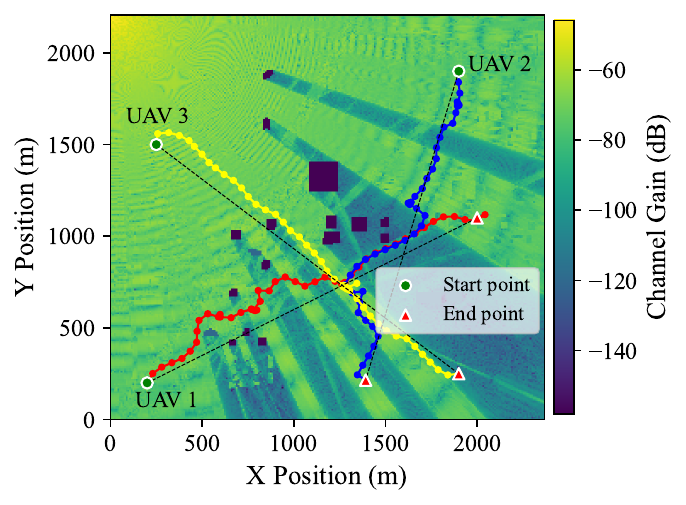}}
		\subfloat[]{\includegraphics[width=0.33\columnwidth]{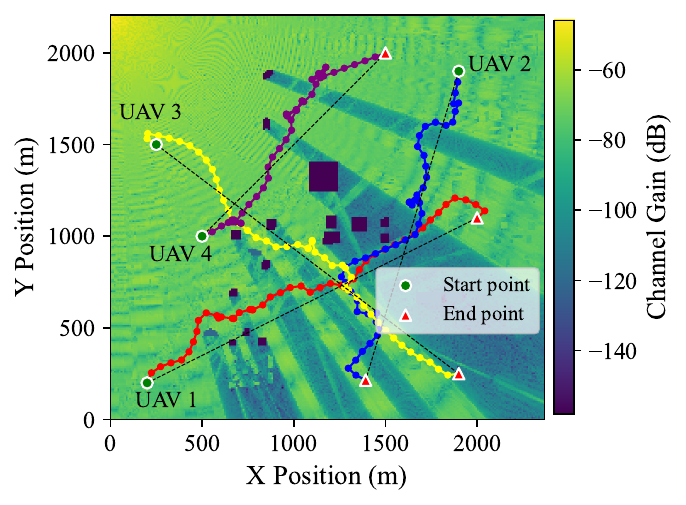}}
		\vspace{-0.16in}
		\\
		\subfloat[]{\includegraphics[width=0.33\columnwidth]{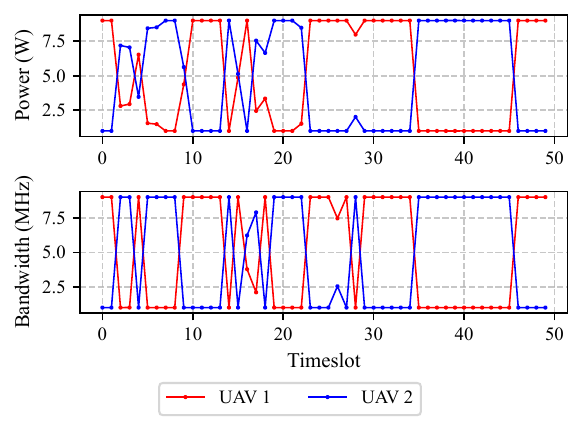}}
		\subfloat[]{\includegraphics[width=0.33\columnwidth]{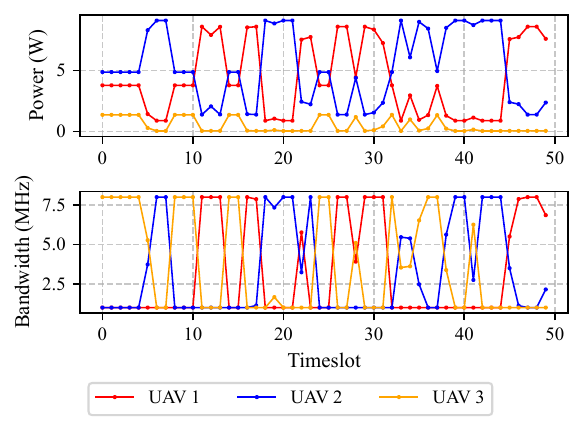}}
		\subfloat[]{\includegraphics[width=0.33\columnwidth]{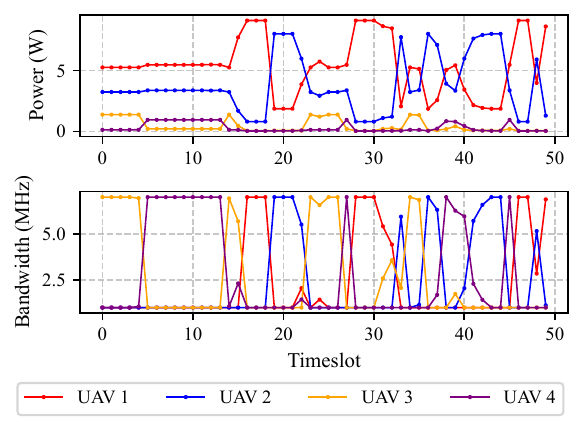}}
		\caption{The trajectories, power, bandwidth, and communication rate optimized by the proposed CKM-JPBTO under cKAN based predicted CKM under channel gain measurements sampling ratio ${\left\vert \mathcal{D}^s \right\vert}/{\vert \mathcal{\tilde{D}} \vert}=3\%$ for (a)(d) \textbf{two}-UAV, (b)(e) \textbf{three}-UAV, and (c)(f) \textbf{four}-UAV system with $P_{\max}=10$ W, $B_{\max}=100$ MHz, $N=50$, and $T=100$ s, and the groud truth CKM visualization.}
		\label{fig:trajectories_CKM}
		\vspace{-0.1in}
	\end{figure}
	
	\begin{figure}[!t]
		\centering
		\vspace*{-0.1in}
		\subfloat[]{\includegraphics[width=0.33\columnwidth]{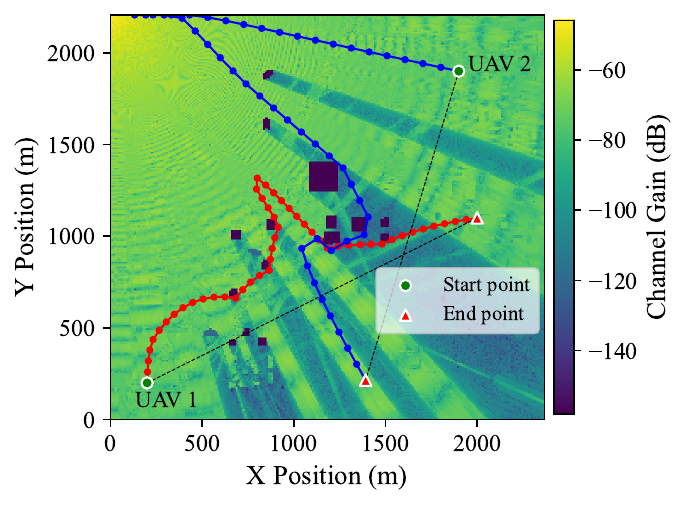}}
		\subfloat[]{\includegraphics[width=0.33\columnwidth]{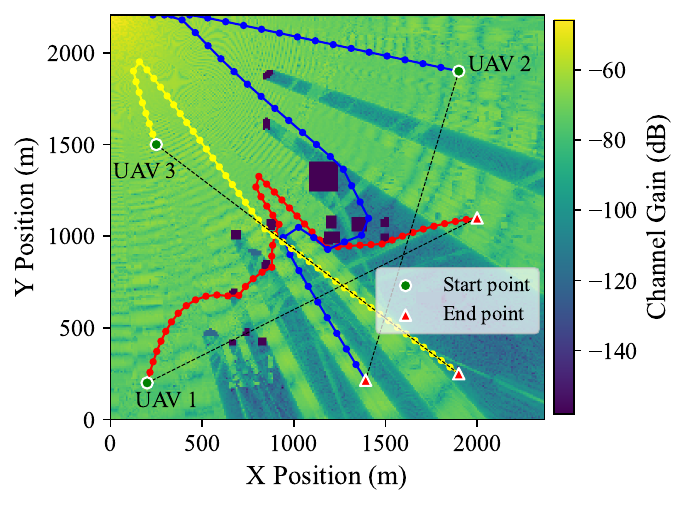}}
		\subfloat[]{\includegraphics[width=0.33\columnwidth]{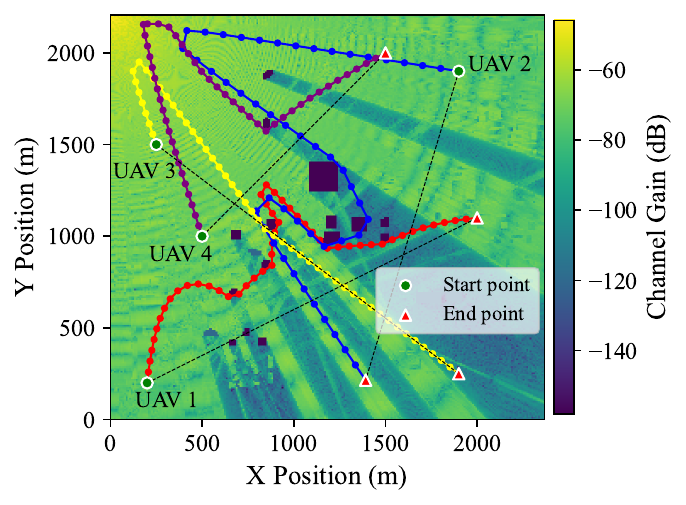}}
		\vspace{-0.16in}\\
		\subfloat[]{\includegraphics[width=0.33\columnwidth]{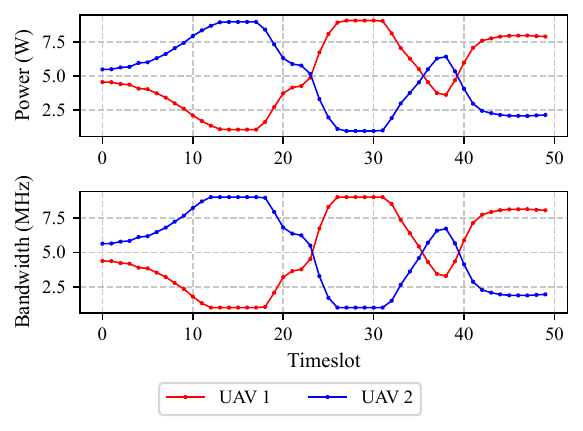}}
		\subfloat[]{\includegraphics[width=0.33\columnwidth]{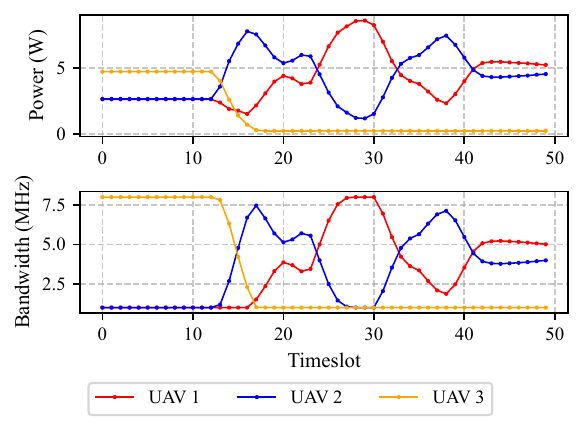}}
		\subfloat[]{\includegraphics[width=0.33\columnwidth]{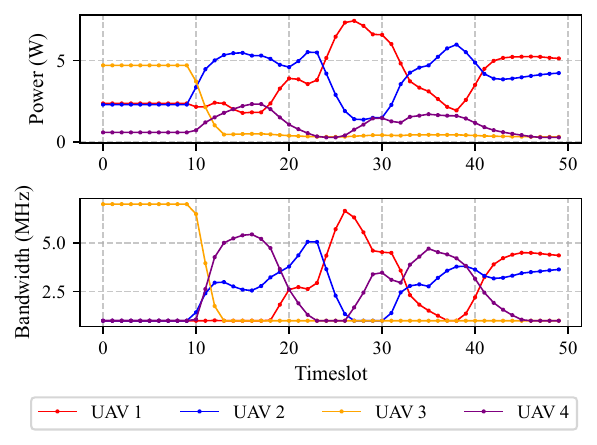}}
		\caption{The trajectories, power, bandwidth, and communication rate optimized by SC-JPBTO for (a)(d) \textbf{two}-UAV, (b)(e) \textbf{three}-UAV, and (c)(f) \textbf{four}-UAV system with $P_{\max}=10$ W, $B_{\max}=100$ MHz, $N=50$, and $T=100$ s, and the groud truth CKM visualization.}
		\label{fig:trajectories_DRL}
		\vspace{-0.1in}
	\end{figure}
	
	\begin{figure}[!t]
		\centering
		\vspace*{-0.1in}
		\subfloat[]{\includegraphics[width=0.33\columnwidth]{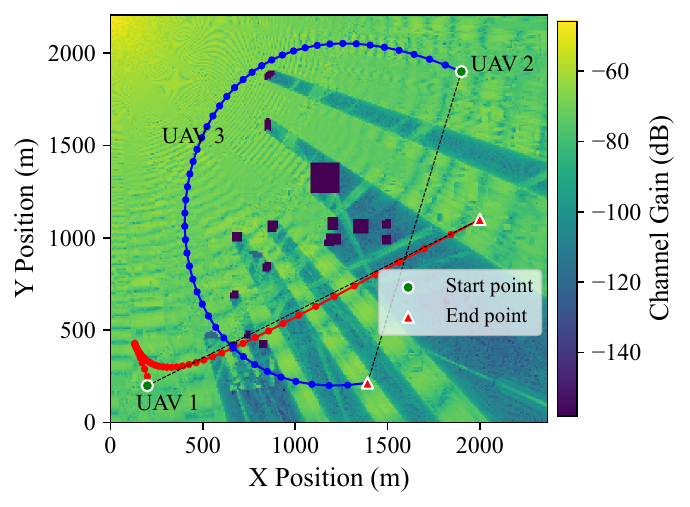}}
		\subfloat[]{\includegraphics[width=0.33\columnwidth]{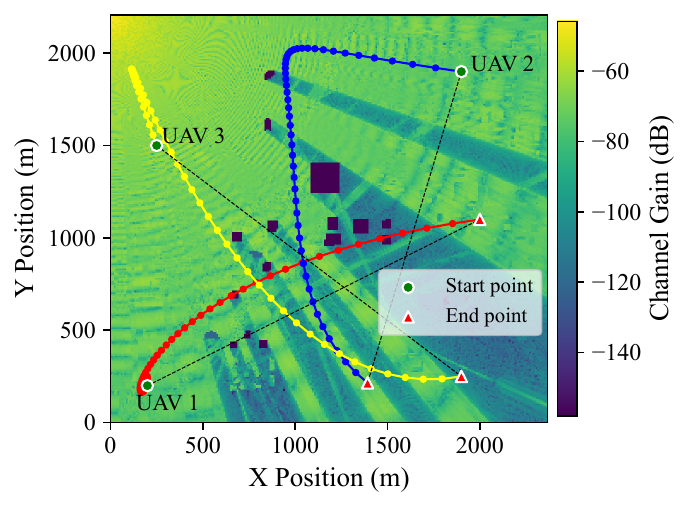}}
		\subfloat[]{\includegraphics[width=0.33\columnwidth]{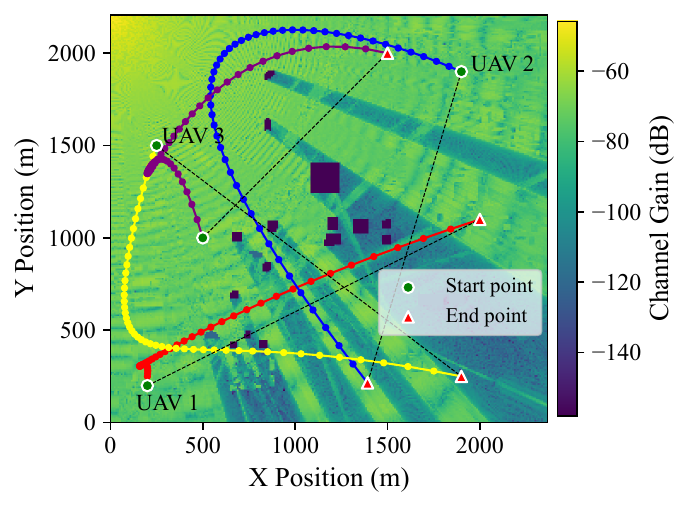}}
		\vspace{-0.16in}\\
		\subfloat[]{\includegraphics[width=0.33\columnwidth]{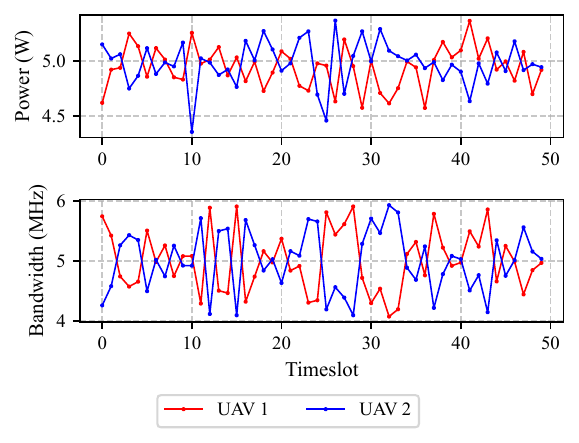}}
		\subfloat[]{\includegraphics[width=0.33\columnwidth]{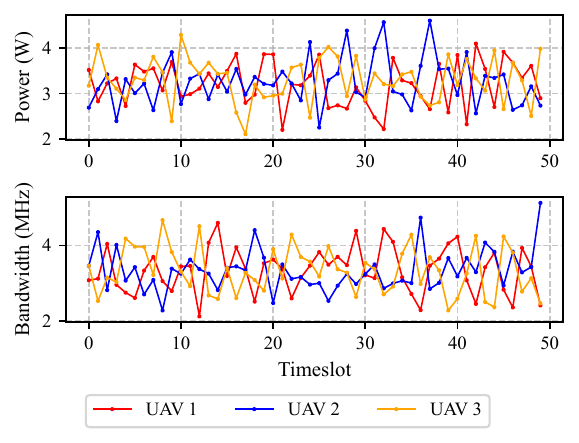}}
		\subfloat[]{\includegraphics[width=0.33\columnwidth]{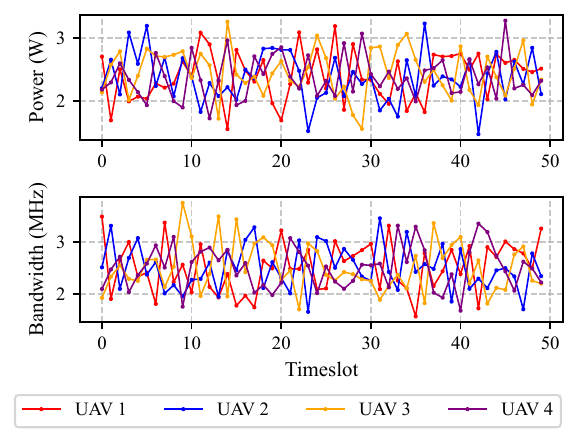}}
		\caption{The trajectories, power, bandwidth, and communication rate optimized by GWO-JPBTO for (a)(d) \textbf{two}-UAV, (b)(e) \textbf{three}-UAV, and (c)(f) \textbf{four}-UAV system with $P_{\max}=10$ W, $B_{\max}=100$ MHz, $N=50$, and $T=100$ s, and the groud truth CKM visualization.}
		\label{fig:trajectories_LoS}
		\vspace{-0.2in}
	\end{figure}

	\vspace{-0.1in}
	\subsection{Performance of CKM-JPBTO}
	We now evaluate the trajectory design performance based on the constructed CKM. For benchmarking, we consider the following two baselines:
	\begin{enumerate}
		\item[-] SC-JPBTO: the same AO framework but with a distance-dependent statistical channel (SC) model instead of the CKM. In this baseline, the channel gain is approximated as $h_m[n] = \beta_0 / d_m[n]^2$, where $d_m[n] = \Vert \mathbf{q}_m[n] - \mathbf{q}_{\rm BS} \Vert_2$ is the Euclidean distance to the BS, and $\beta_0$ is the channel gain at the reference distance of 1 meter. This simplifies the rate to $R_m[n] = \alpha_m[n] B_{\max} \log_2\left(1 + \frac{p_m[n] \beta_0}{N_0 \alpha_m[n] B_{\max} d_m[n]^2}\right)$, ignoring site-specific effects of NLoS by buildings.
		\item[-] GWO-JPBTO: jointly optimizing the power, bandwidth, and trajectory of UAVs via non gradiant GWO \cite{MIRJALILI201446}, and the grid-based CKM constructed by RadioUNet \cite{Levie_2021_TWC}.
	\end{enumerate}

	\myreffig{fig:Convergence} illustrates the convergence behavior of the proposed algorithm. It can be observed that under different UAV numbers $M \in \{2, 3, 4\}$, the objective function in $\rm (P0)$ converges within 15 iterations, demonstrating the algorithm's efficiency. Note that the objective function value in \myreffig{fig:Convergence} is calculated from the predicted channel gain $\hat{\mathcal{H}}(\mathbf{q})$ based on a training set under sampling ratio $\vert \mathcal{D}^s \vert / \vert \tilde{\mathcal{D}} \vert = 3\%$, and is therefore higher than the true rate calculated with the ground-truth CKM\footnote{Note that obtaining the channel gain from the ground-truth CKM maps an arbitrary position $\mathbf{q} \in \mathcal{D}$ to the closest discrete position $\tilde{\mathbf{q}} \in \tilde{\mathcal{D}}$ based on its grid location to acquire a more accurate channel gain, which is then used to compute the communication rate.}.
	The results in \myreftable{tab:runtime} shows that the proposed cKAN significantly reduces inference latency versus the generative RadioUNet. Furthermore, by leveraging efficient gradient computation (0.088 s), the proposed CKM-JPBTO achieves a per-iteration time of only 0.293 s—approximately 3× faster than the heuristic GWO-JPBTO. This efficiency enables full trajectory design within seconds, supporting quasi-real-time applications.

	Figs. \ref{fig:trajectories_CKM}-\ref{fig:trajectories_LoS} illustrate the optimized trajectories alongside power and bandwidth allocation results for CKM-JPBTO, SC-JPBTO, and GWO-JPBTO, respectively. By leveraging the differentiable CKM, CKM-JPBTO yields trajectories that precisely traverse high-gain reflection regions. In contrast, SC-JPBTO tends to approach the BS directly without avoiding NLoS areas. Furthermore, the non-gradient-based GWO-JPBTO exhibits poor convergence, failing to effectively support the joint optimization of the highly coupled variables.

\begin{figure}[!t]
		\centering
		\includegraphics[width=0.725\columnwidth]{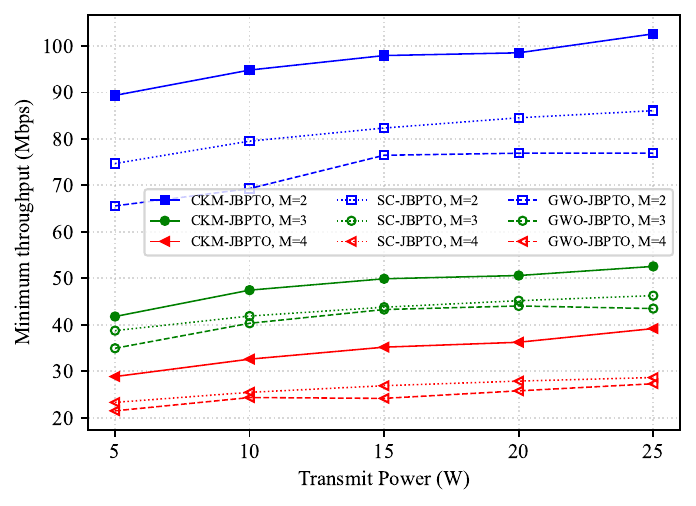}
		\caption{Minimum achievable rate under different transmit power $P_{\max}$ and $M \in \{2,3,4\}$, with $B_{\max} = 10$ MHz, $N=50$, and $T=100$ s.}
		\label{fig:Rate_vs_Power}
	\end{figure}
\begin{figure}[!t]
		\centering
		\includegraphics[width=0.725\columnwidth]{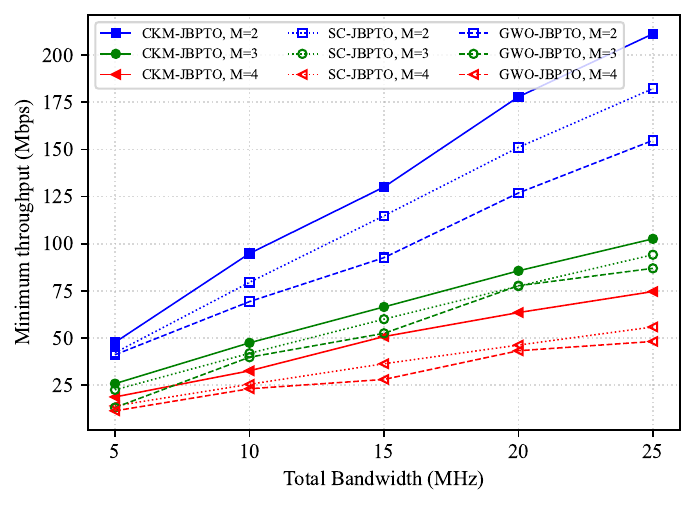}
		\caption{Minimum achievable rate under different bandwidth $B_{\max}$ and $M \in \{2,3,4\}$, with $P_{\max} = 10$ W, $N=50$, and $T=100$ s.}
		\label{fig:Rate_vs_Bandwidth}
	\end{figure}
\begin{figure}[!t]
		\centering
		\includegraphics[width=0.725\columnwidth]{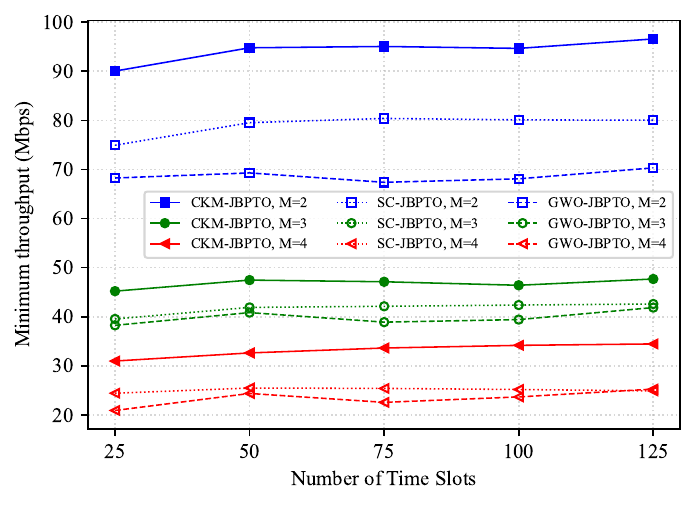}
		\caption{Minimum achievable rate under different timeslot number $N$ and $M \in \{2,3,4\}$, with $P_{\max} = 10$ W, $B_{\max} = 10$ MHz, and $T=100$ s.}
		\label{fig:Rate_vs_TimeslotNum}
	\end{figure}

	In \myreffig{fig:Rate_vs_Power}, \myreffig{fig:Rate_vs_Bandwidth}, and \myreffig{fig:Rate_vs_TimeslotNum}, we present the minimum throughput performance versus $P_{\max}$, $B_{\max}$, and $N$, as the number of UAVs varies with $M \in \{2, 3, 4\}$. These curves represent the averaged results over 20 Monte Carlo trials to eliminate the randomness of initialization. Specifically, \myreffig{fig:Rate_vs_Power} and \myreffig{fig:Rate_vs_Bandwidth} validate the impact of resource constraints. The results indicate that increasing power or bandwidth improves the minimum rate across all configurations.
	It is observed that GWO-JPBTO consistently yields the lowest rate. This highlights the difficulty of optimizing high‑dimensional trajectories ($M$ UAVs $\times$ $N$ time slots $\times$ $N_{W_1}\cdot N_{W_2}$ coordinates) with gradient‑free methods. Without gradient guidance, such approaches become computationally prohibitive and often converge to clearly suboptimal solutions.
	The near-linear growth in \myreffig{fig:Rate_vs_Bandwidth} for CKM-JPBTO confirms the accuracy and effectiveness of the constructed differentiable CKM. However, as power operates inside the logarithmic term in \eqref{eq:rate_expression}, its effect is less pronounced compared to bandwidth. In \myreffig{fig:Rate_vs_TimeslotNum}, as the number of time slots $N$ increases, CKM-JPBTO shows a steady performance gain due to finer trajectory discretization. In contrast, SC-JPBTO and GWO-JPBTO remain relatively stagnant, bounded by their inability to exploit environment-specific channel variations.

%

	\section{Conclusion}
	In this paper, we proposed a novel differentiable CKM-triggered deployment framework for multi-UAV systems. We first introduced a CNN encoder to extract environmental topology features, which could be integrated with both MLP and KAN regressors. This design enhanced conventional coordinate-based regression and further led to cMLP/cKAN that effectively fused spatial features with continuous location inputs. The resulting cMLP/cKAN-based CKM maintained well differentiability with respect to UAV locations, enabling seamless integration with gradient-based optimizers. For LAWN deployment, we developed CKM-JPBTO, a joint power, bandwidth, and trajectory optimization method that leveraged the differentiable CKM within an alternating optimization framework. Simulation results demonstrated that the proposed cKAN-based CKM achieved superior accuracy in modeling site-specific channel propagation gain. And the CKM-JPBTO substantially enhanced the minimum throughput compared to the conventional statistical channel model-based JPBTO. Furthermore, we identified a CKM accuracy threshold required to ensure significant performance improvement, which highlights the minimum quality requirement for CKM construction, guiding optimizing data collection efforts in practical LAWN deployment.


	\bibliographystyle{IEEEtran}
	\bibliography{IEEEabrv, Ref} 

\end{document}